\documentclass[10pt,journal,compsoc]{IEEEtran}
\usepackage{amsmath,amsfonts}
\usepackage{graphicx}
\usepackage[noend,linesnumbered, ruled]{algorithm2e}
\usepackage{array}
\usepackage[caption=false,font=footnotesize,labelfont=rm,textfont=rm]{subfig}
\usepackage{textcomp}
\usepackage{stfloats}
\usepackage{url}
\usepackage{verbatim}

\usepackage{cite}

\usepackage{amssymb}
\usepackage{color}
\usepackage{multirow}
\usepackage{tabularx}
\usepackage{amssymb}
\usepackage{pifont}
\usepackage{booktabs} 
\begin{document}

\title{Secure Low-altitude Maritime Communications via Intelligent Jamming\\
}

\author{Jiawei Huang,
        Aimin Wang,
        Geng Sun,~\IEEEmembership{Senior Member,~IEEE,}
    Jiahui~Li, 
    Jiacheng Wang, \\
    Weijie Yuan,~\IEEEmembership{Senior Member,~IEEE,}
    Dusit Niyato,~\IEEEmembership{Fellow,~IEEE,}
    Xianbin Wang,~\IEEEmembership{Fellow,~IEEE}
\thanks{
\indent Jiawei Huang, Aimin Wang, and Jiahui Li are with the College of Computer Science and Technology, Jilin University, Changchun 130012, China, and Key Laboratory of Symbolic Computation and Knowledge Engineering of Ministry of Education, Jilin University, Changchun 130012, China (E-mails: huangjiawei97@foxmail.com, wangam@jlu.edu.cn, lijiahui@jlu.edu.cn).\\
\indent Geng Sun is with the College of Computer Science and Technology, Key Laboratory of Symbolic Computation and Knowledge Engineering of Ministry of Education, Jilin University, Changchun 130012, China, and also with the College of Computing and Data Science, Nanyang Technological University, Singapore 639798 (E-mail: sungeng@jlu.edu.cn).\\
\indent Jiacheng Wang and Dusit Niyato are with the College of Computing and Data Science, Nanyang Technological University, Singapore 639798 (E-mails: jiacheng.wang@ntu.edu.sg, dniyato@ntu.edu.sg).\\
\indent Weijie Yuan is with the School of Automation and Intelligent Manufacturing, Southern University of Science and Technology, Shenzhen 518055, China (Email: yuanwj@sustech.edu.cn).\\
\indent Xianbin Wang is with the Department of Electrical and Computer Engineering, Western University, London ON N6A 3K7, Canada (E-mail: xianbin.wang@uwo.ca).\\
\indent (Corresponding authors: Geng Sun and Jiahui Li)
}
}

\IEEEtitleabstractindextext{
\begin{abstract}
Low-altitude wireless networks (LAWNs) have emerged as a viable solution for maritime communications. In these maritime LAWNs, unmanned aerial vehicles (UAVs) serve as practical low-altitude platforms for wireless communications due to their flexibility and ease of deployment. However, the open and clear UAV communication channels make maritime LAWNs vulnerable to eavesdropping attacks. Existing security approaches often assume eavesdroppers follow predefined trajectories, which fails to capture the dynamic movement patterns of eavesdroppers in realistic maritime environments. To address this challenge, we consider a low-altitude maritime communication system that employs intelligent jamming to counter dynamic eavesdroppers with uncertain positioning to enhance the physical layer security. Since such a system requires balancing the conflicting performance metrics of the secrecy rate and energy consumption of UAVs, we formulate a secure and energy-efficient maritime communication multi-objective optimization problem (SEMCMOP). To solve this dynamic and long-term optimization problem, we first reformulate it as a partially observable Markov decision process (POMDP). We then propose a novel soft actor-critic with conditional variational autoencoder (SAC-CVAE) algorithm, which is a deep reinforcement learning algorithm improved by generative artificial intelligence. Specifically, the SAC-CVAE algorithm employs advantage-conditioned latent representations to disentangle and optimize policies, while enhancing computational efficiency by reducing the state space dimension. Simulation results demonstrate that our proposed intelligent jamming approach achieves secure and energy-efficient maritime communications. Furthermore, comparison results show that the proposed SAC-CVAE algorithm outperforms baseline methods across various eavesdropper movement patterns, simultaneously maximizing the secrecy rate and minimizing the energy consumption of UAVs.
\end{abstract}

\begin{IEEEkeywords}
Low-altitude maritime communications, physical layer security, UAV-assisted, multi-objective optimization, deep reinforcement learning.
\end{IEEEkeywords}}

%
\maketitle

\IEEEdisplaynontitleabstractindextext

\IEEEpeerreviewmaketitle

\IEEEraisesectionheading{\section{Introduction}
\label{sec:introduction}}

\par \IEEEPARstart{T}{he} expansion of maritime activities has intensified the demand for reliable communication systems to support offshore operations, navigation safety, and emergency responses~\cite{Cao2025}. Nevertheless, the deployment of terrestrial infrastructure in maritime environments faces significant challenges due to harsh marine conditions~\cite{Xu2025}. Consequently, various auxiliary platforms are deployed to facilitate maritime communications. For instance, satellites can provide wide-area coverage, enabling effective data exchange for vessels at sea~\cite{Hu2024}. Yet, satellites often suffer from significant propagation latency due to their long transmission distances. Meanwhile, low-altitude wireless networks (LAWNs) have demonstrated promising potential in maritime communications~\cite{Zhang2025a}. Specifically, rapidly deployable and highly mobile unmanned aerial vehicles (UAVs) are well-suited for maritime LAWNs to enable on-demand coverage~\cite{Zhou2025}. Unfortunately, the open and clear wireless channels of UAVs can be vulnerable to eavesdropping by malicious users, posing security risks. Although conventional cryptography methods can mitigate security threats in certain scenarios~\cite{Jeon2025}, their complex key distribution and management processes introduce communication latency. Particularly, when handling large-scale data transmissions, these methods impose additional burdens on resource-constrained maritime environments.

\par In this case, physical layer security (PLS) can be a promising alternative, which provides an adjustable mechanism through strategic power allocation and artificial noise distribution, thereby enabling adaptation to diverse communication security requirements~\cite{Khoshafa2025}. For example, UAVs, as mobile jammers, enable dynamic coverage adaptation through rapid deployment and position optimization~\cite{Dai2025}, enabling transmission of friendly-jamming signals to enhance PLS. Likewise, the authors in~\cite{Wang2021a} investigated a UAV-enabled secure communication system in which a UAV transmits artificial noise signals to confuse malicious eavesdroppers. Moreover, the authors in~\cite{Huang2025} considered a dual UAV cluster-assisted secure maritime communication system, in which one cluster transmits data signals while the other jams a remote eavesdropper. Nevertheless, these works considered static marine eavesdroppers, while neglecting their dynamic positional variations. The authors in~\cite{Wu2023} considered mobile eavesdroppers and used cooperative UAVs to regulate transmission rates, which enhances the secure performance of communication environments. However, this work assumes that the trajectories of eavesdroppers are predetermined, which may invalidate the method when the real-time eavesdropper positions are not known, thus potentially compromising the security mechanism. This leads us to further consider the need to estimate the dynamic and uncertain positions of eavesdroppers in UAV-assisted maritime PLS communication systems.

\par The implementation of such a system faces several critical challenges. \textit{First}, the mobility of vessels and eavesdroppers induces dynamic wireless channel conditions, while uncertain eavesdropper trajectories complicate the system, rendering conventional offline optimization approaches (\textit{i.e.}, convex optimization and evolutionary computation) ineffective under time-varying maritime scenarios~\cite{Wang2023},~\cite{Yuan2020}. \textit{Second}, we need to precisely control the 3D positions and transmit powers of cooperative UAVs to ensure system security. Such frequent position adjustments of UAVs significantly increase their energy consumption, which poses a trade-off between security performance and energy efficiency. Thus, traditional single-objective optimization frameworks (\textit{e.g.}, ~\cite{Zhang2025},~\cite{Ning2025}) are insufficient, necessitating a novel approach to capture this trade-off. \textit{Finally}, our considered system focuses on a long-term trajectory optimization process that inherently demands precise characterization of multi-modal decision spaces, which further adds to system complexity. Therefore, an innovative approach is required to address dynamic uncertainties, multi-objective trade-offs, and multi-modal optimization challenges in the maritime PLS communication system.

\par To overcome these challenges, we formulate a multi-objective optimization problem (MOP) and propose a generative AI (GenAI)-improved deep reinforcement learning (DRL) algorithm. Our primary contributions are summarized as follows.
\begin{itemize}
    \item \textbf{\textit{Intelligent Jamming for Low-altitude Maritime Communication System:}} We consider a low-altitude maritime communication system with dynamic and uncertain eavesdropper positioning, in which one UAV, as a relay, sends data signals to a marine vessel, and the other UAV, as a jammer, intelligently sends jamming signals to an eavesdropper. To the best of our knowledge, this work is the first to consider dynamic and uncertain eavesdropper trajectories and to design an intelligent jamming mechanism for real-time maritime secure communications.
    
    \item \textbf{\textit{Dynamic and Long-term Multi-objective Optimization Problem Formulation:}} In the low-altitude maritime communication system, security performance and energy efficiency conflict with each other, exhibiting inherent trade-offs. In this case, we formulate a secure and energy-efficient maritime communication MOP (SEMCMOP) that simultaneously maximizes the secrecy rate and minimizes the energy consumption of UAVs. The SEMCMOP accounts for the sequential decision-making process of UAVs across time slots. Consequently, this dynamic and long-term problem requires balancing immediate and sustainable performance rewards throughout the mission duration, which further complicates this problem.

     \item \textbf{\textit{Improved DRL Algorithm by Incorporating GenAI:}} Given the NP-hard complexity and dynamics of the formulated SEMCMOP, we propose a novel soft actor-critic with conditional variational autoencoder (SAC-CVAE) algorithm, which incorporates GenAI capabilities to solve the problem. Specifically, we first transform the problem into a partially observable Markov decision process (POMDP). Then, the SAC-CVAE algorithm can disentangle and optimize policies through an advantage-conditioned latent representation while enhancing computational efficiency by reducing the state space dimension via a long short-term memory (LSTM)-assisted prediction mechanism.

    \item \textbf{\textit{Performance Evaluations and Analyses:}} Simulation results demonstrate that the proposed intelligent jamming approach can achieve secure and energy-efficient maritime communications. Moreover, a comparative analysis with the non-jamming approach confirms the effectiveness of our UAV intelligent jamming approach. In addition, comparison results further show that our proposed SAC-CVAE algorithm outperforms other conventional DRL algorithms across various eavesdropper movement patterns, further indicating its efficiency and robustness.
\end{itemize}

\par The rest of this paper is structured as follows: Section \ref{sec:Related work} reviews the related work. Section \ref{sec:models_and_formulation} presents the models and preliminaries. Section \ref{sec:problem formulation and analysis} formulates and analyzes the SEMCMOP. The GenAI-improved DRL algorithm is proposed in Section \ref{sec:algorithm}. Section \ref{sec:simulation-results-and-analysis} illustrates the simulation results, and Section \ref{sec:conclusion} concludes the overall work.

%
\section{Related work} 
\label{sec:Related work}

\par In this section, we present a review of relevant work related to UAV-assisted maritime communications, security mechanisms for maritime networks, and optimization approaches.

\subsection{UAV-assisted Maritime Communications}

\par Rapidly expanding maritime activities necessitate increasingly reliable maritime communications. Given the high costs and technical challenges of deploying fixed infrastructure across vast ocean areas, auxiliary platforms serve as practical alternatives~\cite{Vangala2024}. For example, the authors in~\cite{Wu2024_2} developed an intelligent spectrum-sharing scheme for satellite-maritime integrated networks, improving throughput and spectral efficiency. Moreover, the authors in~\cite{Li2025} investigated the distributions of aggregated interference with uplink power control to maximize the probability of effective coverage in satellite-maritime networks. While satellite networks provide wide coverage, they introduce substantial propagation delays that particularly challenge real-time applications. Furthermore, the authors in~\cite{Zeng2022} demonstrated a multi-antenna unmanned surface vehicle (USV) system to maximize sum throughput through cooperative beamforming and optimal trajectory planning. However, sea surface reflections cause multipath effects that deteriorate the quality of received signals, while wave-induced platform oscillations degrade antenna alignment~\cite{Liu2021}.

\par In recent years, LAWNs have emerged as an effective solution for maritime communications~\cite{Zhang2025a}. Specifically, UAVs, with exceptional operational flexibility and rapid deployment capabilities, can be integrated into LAWNs to address the challenges of infrastructure deployment in marine environments~\cite{Yu2025}. For instance, the authors in~\cite{Nomikos2024} incorporated UAVs into maritime communication networks (MCNs) to complement shore base stations with limited coverage, thereby improving wireless connectivity and resource efficiency. In~\cite{Liu2022a}, the authors developed a two-layer UAV-based maritime communication mobile edge computing (MEC) network to minimize latency for both communication and computation. Moreover, the authors in~\cite{Yang2020} utilized UAVs to form a cognitive mobile computing network for cooperative search and rescue at sea, enhancing communication throughput. Note that the open and clear wireless channels of UAVs make them extremely vulnerable to eavesdropping attacks during data transmission. However, the aforementioned studies focused on communication efficiency and overlooked this critical security risk.

\subsection{Security Mechanisms of Maritime Networks}

\par To address the security risks in maritime networks, researchers have recently explored various security mechanisms~\cite{Lee2023}. For example, the authors in~\cite{Luo2024} provided a routing protocol to enhance inter-UAV communication efficiency while introducing digital twin technology to guarantee network security. Moreover, the authors in~\cite{Min2025} presented a federated privacy-preserving framework for UAV data collection to optimize autonomous path planning and protect sensitive maritime information. However, these cryptography methods require significant computational resources when processing large-scale data. The resulting power demands and transmission latency make such methods inadequate for maritime missions requiring real-time communications.

\par In this case, the PLS mechanism enables dynamic adjustment of protective measures in response to channel characteristics, thereby ensuring reliable and secure maritime communications~\cite{Wang2025}. Meanwhile, highly mobile and flexible UAVs can serve as platforms of friendly-jamming for security enhancement. For instance, the authors in~\cite{Huang2024} utilized UAVs to form a maritime UAV-enabled virtual antenna array that transmits jamming signals to achieve PLS for vessel communications. However, this work considered static eavesdroppers, limiting its applicability in real-world maritime scenarios where threats are typically mobile. Furthermore, the authors in~\cite{Lu2024} proposed an efficient communication scheme for UAV-relay-assisted maritime MEC with a moving eavesdropper to maximize the secure computing capacity. In addition, the authors in~\cite{Yang2024} investigated a UAV-reconfigurable intelligent surface (RIS)-assisted maritime communication system, maximizing energy efficiency while guaranteeing the quality of service requirements against jamming attacks. Nevertheless, the aforementioned works have a common limitation of assuming predefined eavesdropper trajectories. This assumption disregards the randomness and adaptability of eavesdropper movements, causing performance degradation when deploying the trained model in real-world scenarios.

\subsection{Optimization Approaches}

\par Several approaches have been proposed to achieve UAV-assisted secure maritime communications. For instance, the authors in~\cite{Liu2021} investigated a dual-UAV secure communication system with imperfect eavesdropper location information, while employing a problem decomposition methodology to optimize UAV parameters. However, this method leads to suboptimal solutions, as it treats interconnected system parameters independently. Furthermore, DRL algorithms are common and effective methods for dealing with dynamic optimization problems~\cite{Sun2025}. In~\cite{Lin2024}, the authors proposed an MCN with aerial RIS-assisted UAVs against jamming, while designing a novel penalized DRL algorithm to maximize energy efficiency. Moreover, the authors in~\cite{Liu2023} proposed a UAV relay policy based on reinforcement learning for maritime communications to resist jamming attacks and reduce energy consumption. However, the aforementioned works often treated energy as a constraint and overlooked the complex trade-offs between security and energy consumption, which makes it difficult to obtain appropriate solutions under different energy priority conditions.

\par To achieve the complex trade-off among competing considerations, the multi-objective optimization problem (MOP) framework offers a mathematical foundation to simultaneously optimize multiple conflicting objectives. This framework enables systematic modeling of objective relationships and identification of optimal compromise solutions under varying conditions~\cite{Karami2022}. For example, the authors in~\cite{Sun2023} considered a UAV-enabled secure communication system and formulated an MOP to maximize the worst-case secrecy rate and minimize the energy consumption of UAVs to achieve trade-offs. In addition, the authors in~\cite{Sun2024a} considered a multi-UAV-assisted MEC system and formulated an MOP to meet the computation-intensive and delay-sensitive demands of users. Note that dynamic maritime communication systems require real-time responses to changing wireless channel conditions to ensure reliable signal transmission. However, the aforementioned works employed evolutionary computation methods, which exhibit excessive computational latency and are inadequate for real-time adaptation in such scenarios.

\subsection{Summary}

\par Different from previous works, we consider an intelligent jamming scheme for a low-altitude maritime communication system with dynamic and uncertain eavesdropper positioning. Accordingly, we propose a novel approach to solve the dynamic optimization problem that requires balancing multiple objectives.

%
\section{Models and preliminaries} \label{sec:models_and_formulation}

\par In this section, we first consider the low-altitude maritime communication system with dynamic and uncertain eavesdropper positioning. Then, we detail the vessel movement model. Subsequently, we present the corresponding communication model. Finally, the energy consumption model of the UAV is introduced. Note that the main notations are shown in Table \ref{table:notations}.

\begin{table}[!t]
\renewcommand{\arraystretch}{1.1}
\newcommand{\tabincell}[2]{\begin{tabular}{@{}#1@{}}#2\end{tabular}}
\caption{Main notations}
\label{table:notations}
\centering
\begin{tabular}{ll}
\noalign{\hrule height 1pt}
\bfseries Notation & \bfseries Definition\\
\hline 
& Notations in the system model\\
\hline
$\boldsymbol{A}_{m}$ & Added mass matrix \\
$a_r$ & Area of the rotor disks \\
$\beta_{U,V}$ & Path loss of the U2V link \\
$\beta_{U,U}$ & Path loss of the U2U link \\
$c_d$ & Drag coefficient of the airframe \\
$\mathcal{C}_{U,V}$ & Composite channel of the U2V link \\
$\boldsymbol{C}(\boldsymbol{\upsilon})$ &  Coriolis coefficient matrix \\
$d_{U,V}$ & Distance between the UAV and vessel \\
$d_{U,U}$ & Distance between UAVs \\
$\boldsymbol{D}(\boldsymbol{\upsilon})$ &  Damping coefficient matrix \\
$P_{A}$ & Transmit power of Alice \\
$P_{B}$ & Transmit power of Bob \\
$r_r$ & Rotor solidity \\
$\boldsymbol{R}_{m}$ & Rigid-body mass matrix \\
$s_m$ & Mean induced flow speed  \\ 
$s_r$ & Tip speed of the rotating blades\\
$v_h$ & UAV horizontal velocity \\
$v_v$ & UAV vertical velocity \\
$v_f$ & UAV forward velocity \\
$\rho$ & Atmospheric density \\
$\varsigma $ & Gaussian random variable of the U2V link \\
\hline
& Notations in the algorithm\\
\hline
$\mathcal{A}$ & Action space set \\ 
$\alpha$ & Temperature parameter of SAC\\
$\boldsymbol{c}$ & State-advantage condition of the CVAE \\
$\mathcal{D}$ & Replay buffer \\ 
$\imath$ & Coefficient for the KL-divergence loss term \\
$LSTM_f$ & Forget gate of the LSTM network \\
$LSTM_i$ & Input gate of the LSTM network \\
$LSTM_o$ & Output gate of the LSTM network \\
$\mathcal{O}$ & Observation space set \\
$\omega_{m}$ & Weights for the optimization objective $m$ \\
$p_{\delta}$  & Decoder of the CVAE \\
$\pi_{\Phi}$ & Tractable policy \\
$q_{\varphi}$ & Encoder of the CVAE \\
$Q_{\theta}$ & Soft Q-value network \\
$\mathcal{R}$ & Reward value \\
$\mathcal{S}$ & Global state space set\\
$V_{\psi}$ & State-value network\\
$Z$ & Storage length of the historical trajectory sequence \\
$\zeta$ & Advantage value of the CVAE \\
$\zeta^*$ & Maximum advantage value of the CVAE \\
$\boldsymbol{z}$ & Latent representation of the CVAE \\
\noalign{\hrule height 1pt}
\end{tabular}
\end{table}

\subsection{System Overview}

\par
\begin{figure}
\centering
\includegraphics[width=3.5in]{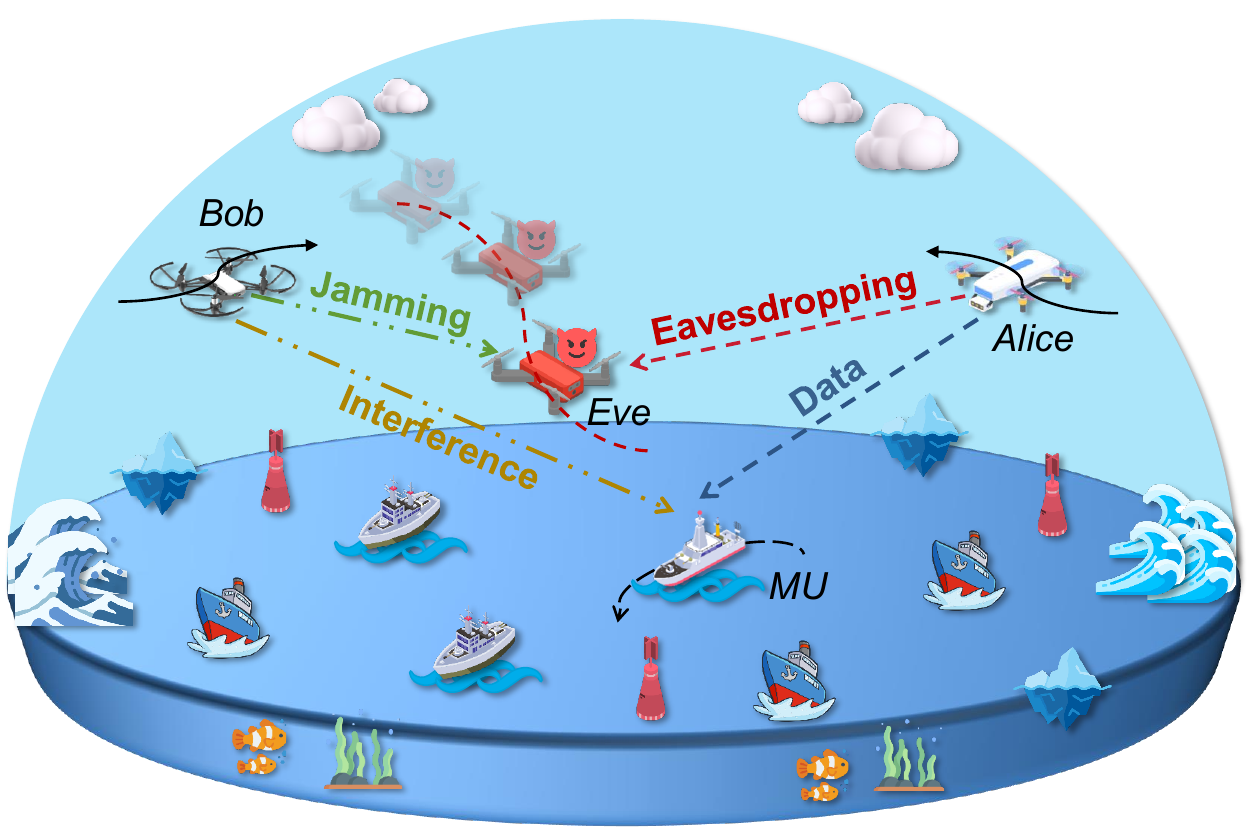}
\caption{A low-altitude maritime communication system with dynamic and uncertain eavesdropper positioning.}
\label{fig:scenario-model}
\end{figure}

 As shown in Fig. \ref{fig:scenario-model}, we consider a low-altitude maritime communication system with dynamic and uncertain eavesdropper position, which includes a marine user (MU), a legitimate UAV denoted as Alice, an illegitimate UAV denoted as Eve, and an assisted UAV denoted as Bob. Specifically, an MU may not be able to receive signals from base stations or other long-range users due to the challenges in deploying infrastructure at sea~\cite{Yang2024}. In such cases, UAVs, with their high mobility and flexibility, serve as efficient low-altitude platforms to forward data to MUs. However, the open and clear channels of UAVs make data signals susceptible to eavesdropping by Eve, whose position is dynamic and uncertain. In this case, another mobile UAV at sea, i.e., Bob, can act as a friendly jammer to jam Eve, so that ensuring data security and integrity.

\par Without loss of generality, we divide the total serving time $T$ into $N$ time slots with equal duration $d_{n}=T/N$, denoted by the set $\mathbb{N}  \triangleq \left \{ 1,2,\dots, N \right \}$. An MU follows its navigation trajectory, and Alice sends signals through the data link. When Eve attempts to obtain data from Alice via the eavesdropping link, Bob sends jamming signals to Eve to degrade the eavesdropping channel and ensure the security of the data signals. Additionally, the MU navigates along a specified route. However, due to the dynamic and uncertain flight path of Eve, its complete position cannot be obtained in advance. Thus, it is crucial to determine the position of Eve to ensure the effectiveness of the jamming strategy.

\par In the communication process, we use the three-dimensional (3D) Cartesian coordinate system to represent the time-varying locations of the MU, Alice, Eve, and Bob at time slot $n$ as ($x_{M}[n]$, $y_{M}[n]$, $z_{M}[n]$), ($x_{A}[n]$, $y_{A}[n]$, $z_{A}[n]$), ($x_{E}[n]$, $y_{E}[n]$, $z_{E}[n]$), and ($x_{B}[n]$, $y_{B}[n]$, $z_{B}[n]$), respectively. Note that the jamming signals may interfere with an MU, affecting the effective reception of the data. To evaluate this interference and optimize system performance, we next provide the vessel movement model and communication model.

\subsection{Vessel Movement Model}

\par The movement of a vessel is often described by using two 3D right-handed Cartesian coordinate systems~\cite{Ren2023}. The first is a normal coordinate system, denoted as $\mathfrak{n}$, where the origin is placed on the sea surface, $x$, $y$, and $z$ axes are aligned with the north, east, and downward directions, respectively. The second system, denoted as $\mathfrak{g}$, is fixed relative to the vessel, with the origin at the center of gravity, and $x^{\mathfrak{g}}$, $y^{\mathfrak{g}}$, and $z^{\mathfrak{g}}$ axes pointing toward the bow, starboard, and downward, respectively. Moreover, rotation around the $x^{\mathfrak{g}}$, $y^{\mathfrak{g}}$, and $z^{\mathfrak{g}}$ axes corresponds to the roll ($\phi$), pitch ($\theta$), and yaw ($\psi$) of the vessel. The rotations are represented by the Euler angle vector $\boldsymbol{\Omega } = [\phi, \theta, \psi]^T$. In addition, the movement of the vessel is modeled by using a six-degree-of-freedom system $\langle x, y, z, \phi, \theta, \psi \rangle$, which mathematically captures the spatial dynamics, and it is expressed by~\cite{Skulstad2020}
\begin{equation}
    \dot{ \boldsymbol{\Upsilon}}[n]=\boldsymbol{\Gamma} (\boldsymbol{\Omega }[n]) \boldsymbol{\upsilon }[n],
    \label{vessel-location1}
\end{equation}
\noindent where the vector $\boldsymbol{\Upsilon}[n] = [x[n], y[n], z[n], \phi[n], \theta[n], \psi[n]]^T$ is the displacement and rotational state at time slot $n$, and $\boldsymbol{\upsilon }[n] = [v_x[n], v_y[n], v_z[n], v_{\phi}[n], v_{\theta}[n], v_{\psi}[n]]^T$ describes the translational and rotational velocities at the same time. Moreover, the derivative of $\boldsymbol{\Upsilon}[n]$, denoted as $\dot{\boldsymbol{\Upsilon}}[n]$, represents the rate of change of the location and orientation, and the matrix $\boldsymbol{\Gamma}$ describes the transformation between the horizontal plane of $\left\{\mathfrak{g}\right\}$ and $\left\{\mathfrak{n}\right\}$. Additionally, the velocity vector is influenced by the following external factors, which can be described by
\begin{equation}
\begin{split}
\left(\boldsymbol{R}_{m}+\boldsymbol{A}_{m}\right) \dot{\boldsymbol{\upsilon }}[n]+&\boldsymbol{C}(\boldsymbol{\upsilon }[n]) \boldsymbol{\upsilon }[n] + \boldsymbol{D}(\boldsymbol{\upsilon }[n]) \boldsymbol{\upsilon }+\boldsymbol{r}(\boldsymbol{\Upsilon})\\ &=\boldsymbol{\iota }_{w}+\boldsymbol{\iota }_{o}+\boldsymbol{\iota }_{wa}+\boldsymbol{\iota }_{t}[n],
\end{split}
\label{vessel-location2}
\end{equation}
\noindent where $\boldsymbol{R}_{m}$ and $\boldsymbol{A}_{m}$ are the rigid-body and added mass matrices, respectively, and $\boldsymbol{C}(\boldsymbol{\upsilon})$ and $\boldsymbol{D}(\boldsymbol{\upsilon })$ denote the Coriolis and damping coefficient matrices, respectively. Moreover, $\dot{\boldsymbol{\upsilon}}[n]$ denotes the time derivative of the velocity vector $\boldsymbol{\upsilon}$, and $\boldsymbol{r}(\boldsymbol{\Upsilon})$ is the resilience. Additionally, the vectors $\boldsymbol{\iota }_w$, $\boldsymbol{\iota }_o$, and $\boldsymbol{\iota }_{wa}$ represent the forces exerted on the vessel by the wind, ocean currents, and waves, respectively, and $\boldsymbol{\iota}_{t}[n]$ corresponds to the thrust generated by the vessel thrusters at time slot $n$. 

\par The spatial relationship between the vessel and UAV influences the effect of data transmission, and dynamic wireless channel conditions impact the signal reception effectiveness. Therefore, we proceed to introduce the communication model.

\subsection{Communication Model}

\par This system focuses on two primary communication links, involving the UAV-to-vessel (U2V) link and UAV-to-UAV (U2U) link. Specifically, the Alice-to-MU data link of the U2V link is used for sending data signals, which could be eavesdropped on by Eve. Meanwhile, the Bob-to-Eve jamming link of the U2U link is designed to jam Eve, potentially interfering with the effective data reception of the MU. The detailed processes are described as follows.

\subsubsection{U2V Link to the MU}

\par The U2V link is established based on channel state information (CSI), which can be obtained from the intended flight trajectories of Alice and Bob, and the precalculated location of the MU. Moreover, since the antenna height at UAVs is much higher than that at vessels, the path loss of the U2V link at time slot $n$ can be calculated by
\begin{equation}
    \beta_{U,V}[n][dB]=10 I_{r} \log_{10}\left(\frac{d_{U,V}[n]}{d_{r}}\right)+\varsigma _{U,V}[n]+P_{d},
\end{equation}
\noindent where $d_{U,V}[n]$ denotes the distance between the UAV and MU at time slot $n$, and $\varsigma_{U,V}[n]$ is a zero-mean Gaussian random variable with standard deviation $\sigma_{X_U}$ at the same time. Moreover, $I_r$ is the relevant index, and $P_d$ denotes the parameter for the reference distance $d_r$. Note that $\beta_{A,M}[n]$ and $\beta_{B,M}[n]$ denote the path losses from Alice and Bob to the MU at time slot $n$, respectively. 

\par Then, the composite channel of the U2V link at time slot $n$ can be denoted as follows:
\begin{equation}
    \mathcal{C}_{U,V}[n] = \sqrt{\beta_{U,V}[n]}
    \left(\sqrt{\frac{F_{V}}{1+F_{V}}}+\sqrt{\frac{1}{1+F_{V}}}{h}_{U,V}[n]\right),
\end{equation}
\noindent where $F_{V}$ indicates the Rician factor, and ${h}_{U,V}[n] \in \mathcal{CN}(0,1)$. Moreover, $\mathcal{C}_{A,M}[n]$ and $\mathcal{C}_{B,M}[n]$ denote the channels from Alice and Bob to the MU at time slot $n$, respectively.

\subsubsection{U2U Link to Eve}

\par Given that the U2U link operates in an aerial environment, its signal propagation follows the free-space path loss model, which is expressed mathematically as follows~\cite{Wang2021}:
\begin{equation}
    \beta_{U,U}[n][dB]=20 \log_{10}^{(d_{U,U}[n])}+20 \log _{10}^{f_c}+20 \log_{10}^{\frac{4\pi}{300}},
\end{equation}
\noindent where $d_{U,U}[n]$ represents the distance between UAVs at time slot $n$ in kilometer (km), and $f_{c}$ is the carrier frequency in MHz. Note that $\beta_{A,E}[n]$ and $\beta_{B,E}[n]$ are the path losses from Alice and Bob to Eve at time slot $n$, respectively. 

\par Based on the U2V and U2U links, the achievable rate of the MU at time slot $n$ can be calculated by
\begin{equation}
    R_{M}[n]=\log_2\left(1+\frac{P_{{A}}[n]G_A \left|\mathcal{C}_{A,M}[n]\right|^{2}}{P_{B}[n]G_B\left|\mathcal{C}_{B,M}[n]\right|^{2}+\sigma^{2}}\right),
    \label{Alice_rate}
\end{equation}
\noindent where $P_{A}[n]$ and $P_{B}[n]$ are the transmit powers of Alice and Bob at time slot $n$, respectively. Moreover, $G_A$ and $G_B$ denote the antenna gains of Alice and Bob, respectively, and $\sigma^{2}$ is the additive white Gaussian noise power.

\par Likewise, the achievable rate of Eve at time slot $n$ is expressed by
\begin{equation}
    R_{E}[n]=\log_2\left(1+\frac{P_{A}[n]G_A \beta_{A,E}[n]}{P_{B}[n]G_B\beta_{B,E}[n]+\sigma^{2}}\right).
    \label{Eve_rate}
\end{equation}

\par Then, we define the immediate secrecy rate of the system, which can be expressed as:
\begin{equation}
    C_s[n]=\left [ R_M[n]-R_E[n] \right ] ^+,
    \label{secrecy}
\end{equation}
\noindent where $C_s[n]$ is non-negative, and $[\chi]^+ \triangleq max(0,\chi)$.

\par Based on the preceding analysis, the controllable 3D locations and transmit powers of Alice and Bob are critical factors to ensure secure maritime communications. During communications, Alice dynamically adapts to the mobile MU, while Bob repositions relative to Eve and MU to optimize jamming effectiveness. These continuous adjustments incur energy costs, requiring careful management for sustained UAV operation. Consequently, we next present the energy consumption model of the UAV.

\subsection{Energy Consumption Model of the UAV}

\par At each time slot $n$, the UAV determines movement by executing a 3D action vector $\mathcal{A}[n] = (A_{x}[n], A_{y}[n], A_{z}[n])$. Specifically, the locational coordinates of the UAV $(x_{U}[n],y_{U}[n],z_{U}[n])$ are subsequently updated by using the displacement increment $\mathcal{A}[n]$, which is derived from the previous location. The iterative process $(x_{U}[n],y_{U}[n],z_{U}[n]) = (x_{U}[n-1],y_{U}[n-1],z_{U}[n-1]) + \mathcal{A}[n]$ governs the trajectory of the UAV.

\par Then, we introduce the energy consumption of the UAV. Specifically, the total energy expenditure of UAVs is categorized into propulsion energy and communication energy. As demonstrated in~\cite{Zeng2018}, the propulsion energy dominates the total energy, while communication-related energy consumption can be negligible by comparison. Consequently, we adopt the propulsion power model for calculating the energy consumption of the UAV during horizontal motion as follows:
\begin{equation}
\begin{split}
    P_U(v_{h}[n]) =& P_{I}\left [ \left( 1+\frac{v_{h}^{4}[n] }{4s_{m}^{4}} \right)^\frac{1}{2} -\frac{v^2_{h}[n]}{2s_{m}^2}  \right ]^{\frac{1}{2}} + P_{p} \\& +\frac{3P_B v^2_{h}[n]}{s^2_{r}}  +\frac{v_{h}^{3}[n]c_{d}r_{r}a_{r}\rho }{2},
\end{split}
\label{p_v}
\end{equation}
\noindent where $v_{h}[n] = \sqrt{(A_{x}[n])^2+(A_{y}[n])^{2}} /d_n$ represents the UAV horizontal velocity at time slot $n$. Moreover, $P_{I}$ and $P_{p}$ are the induced power and blade profile power, respectively, and $s_{r}$ and $s_{m}$ denote the tip speed of the rotating blades and mean induced flow speed through the rotor disk, respectively. In addition, $c_{d}$, $r_{r}$, $a_{r}$, and $\rho$ denote the drag coefficient of the airframe, rotor solidity, area of the rotor disks, and atmospheric density, respectively.

\par Our model excludes energy consumption during the acceleration and deceleration phases of UAVs, as the transient phases constitute a negligible portion of the total operational duration~\cite{Li2024}. Consequently, we employ a simplified approximation model to quantify the energy consumption of UAVs in 3D flying paths, which integrates the propulsion energy for sustained flight, kinetic energy during velocity adjustments, and gravitational energy. The energy consumption of a UAV operating in 3D space is denoted by~\cite{Sun2024a}
\begin{equation}
\begin{split}
E_{U}(N)\approx &\int_{0}^{N} P_U\left(v_{h}[n]\right) dn+\frac{1}{2}m_{U}\left (v_{f}[N]^{2}-v_{f}[0]^{2} \right)\\& +m_{U}g\left(h_U[N]-h_U[0]\right ),
\end{split}
\label{E_t}
\end{equation}
\noindent where $v_{f}[n] = \sqrt{v_{h}^2[n] + v_{v}^2[n]}$ represents the forward velocity of the UAV at time slot $n$, of which $v_{v}[n] = |A_{z}[n]|/d_n$ is the vertical velocity of the UAV at the same time. Additionally, $m_{U}$ and $g$ represent the mass of the UAV and gravitational acceleration, respectively, and $h_U[n]$ denotes the flight height of the UAV at time slot $n$.

%
\section{Problem formulation and analyses}
\label{sec:problem formulation and analysis}

\par In this section, we first present the problem statement, then formulate the optimization problem, and proceed with problem analyses.

\subsection{Problem Statement}

\par Given the challenges of deploying communication infrastructure at sea, the flexible UAV serves as a low-altitude auxiliary platform to facilitate signal transmission to the vessel. However, the signals are vulnerable to eavesdropping by an illegitimate UAV. In this case, another assisted UAV can send jamming signals toward the eavesdropper, thereby degrading the eavesdropping channel and enabling secure maritime communications. However, the jamming signals might interfere with the vessel. To address this issue, we need to precisely control jamming signals, enhancing the effect on the eavesdropper while minimizing interference with the vessel. Therefore, we aim to maximize the secrecy rate of the system.

\par Since the vessel follows its engine-determined routes and executes specific tasks, its location cannot be controlled. Moreover, the location of an external hostile eavesdropper is inherently unmanageable. Therefore, the achievable rates of the MU and Eve are controlled by \textbf{\textit{3D locations}} and \textbf{\textit{transmit powers}} of both Alice and Bob. Note that adjusting the 3D locations of the UAVs leads to increased energy consumption. Thus, minimizing the \textbf{\textit{locational adjustments}} of UAVs is crucial for improving overall energy efficiency.

\par Combining the aforementioned factors, the decision variables to be jointly optimized are the following parameters: \textit{(i)} $\mathbb{L}_A $ = \{$\mathbb{X}_{A}, \mathbb{Y}_{A}, \mathbb{Z}_{A}$\} denotes the 3D location set of Alice over $N$ time slots, where $\mathbb{X}_{A}$ = $ \left \{ x_{A}[n]\right \}_{n=1}^{N}$, $\mathbb{Y}_{A}$ = $ \left \{ y_{A}[n]\right \}_{n=1}^{N}$, and $\mathbb{Z}_{A}$ = $ \left \{ z_{A}[n]\right \}_{n=1}^{N}$. \textit{(ii)}  $\mathbb{P}_A$ = $ \left \{P_{A}[n]\right \}_{n=1}^{N}$ is the transmit power of Alice over $N$ time slots. \textit{(iii)} $\mathbb{L}_B $ = \{$\mathbb{X}_B, \mathbb{Y}_B, \mathbb{Z}_B$\} denotes the 3D location set of Bob over $N$ time slots, where $\mathbb{X}_B$ = $ \left \{ x_{B}[n]\right \}_{n=1}^{N}$, $\mathbb{Y}_B$ = $ \left \{  y_{B}[n]\right \}_{n=1}^{N}$, and $\mathbb{Z}_B$ = $ \left \{ z_{B}[n]\right \}_{n=1}^{N}$. \textit{(iv)} $\mathbb{P}$ = $ \left \{ P_{B}[n]\right \}_{n=1}^{N}$ is the transmit power of Bob over $N$ time slots.

\subsection{Problem Formulation}

\par In our considered system, we focus on the following optimization objectives simultaneously.

\par \textbf{\textit{Optimization Objective 1:}} To achieve secure low-altitude maritime communications, the first optimization objective is to maximize the \textbf{\textit{total secrecy rate}} of the system over $N$ time slots, which is expressed by
\begin{equation}
    f_{1}(\mathbb{L}_{A},\mathbb{P}_{A},\mathbb{L}_{B},\mathbb{P}_{B})= { \sum_{n=1}^{N}} C_{s}[n].
\end{equation}

\par \textbf{\textit{Optimization Objective 2:}} The achievement of the above objective requires frequent adjustments to the positions of Alice and Bob, which consumes their energy. Given the limited energy supply available at sea, the second objective is to minimize the \textbf{\textit{total energy consumption}} of Alice and Bob over $N$ time slots as follows:
\begin{equation}
f_{2}(\mathbb{L}_{A}, \mathbb{L}_{B})= \sum_{n=1}^{N}\left(E_A[n]+E_B[n]\right),
\end{equation}
\noindent where $E_{A}[n]$ and $E_{B}[n]$ are the energy consumptions of Alice and Bob at time slot $n$, respectively.

\par The abovementioned two optimization objectives are conflicting. Specifically, we need to control the positions of Alice and Bob to maximize the secrecy rate of the system, which conflicts with minimizing their energy consumption. Moreover, based on Eq. (\ref{p_v}), higher UAV velocity results in increased energy consumption, while lower velocity prolongs communication time and increases hovering energy consumption. Therefore, the two optimization objectives conflict with each other, necessitating an appropriate modeling method to balance this conflict. In this case, the MOP modeling provides a mathematical framework that simultaneously optimizes multiple conflicting objectives~\cite{Karami2022}, which is well-suited for capturing trade-offs and can be used to formulate our problem.

\par Accordingly, we formulate the SEMCMOP as follows:
\begin{subequations}
\label{con:mop}
\begin{align}
&\min _{\left \{ \mathbb{L}_{A},\mathbb{P}_{A},\mathbb{L}_{B},\mathbb{P}_{B}\right \} } \  F=\left \{ -f_{1},f_{2} \right \}, \label{con:mopZa}\\
\mbox{s.t.}\  
&C1: \mathbb{L}_{Amin}\le \mathbb{L}_{A}[n] \le \mathbb{L}_{Amax}, \forall n\in \mathbb{N},\label{con:mopZb}\\   
&C2: P_{min}\le P_{A}[n] \le P_{max}, \forall n\in \mathbb{N} ,\label{con:mopZc}\\ 
&C3:\mathbb{L}_{Bmin}\le \mathbb{L}_{B}[n] \le \mathbb{L}_{Bmax},  \forall n\in \mathbb{N},\label{con:mopZd} \\
&C4: P_{min}\le P_{B}[n] \le P_{max}, \forall n\in \mathbb{N} ,\label{con:mopZe} \\
&C5: R_M[n] > R_{min}, \forall  n\in \mathbb{N}, \label{con:mopZf} \\
&C6: {\textstyle \sum_{n=1}^{N}} P_{A}[n] \leq P_T, \forall n\in \mathbb{N},\label{con:mopZg} \\
&C7: {\textstyle \sum_{n=1}^{N}} P_{B}[n] \leq P_T, \forall n\in \mathbb{N}, \label{con:mopZh}\\
&C8: P_{B}[n]G_{B}\left|\mathcal{C}_{B,M}[n]\right|^{2} \le I_{0},  \forall n\in \mathbb{N},\label{con:mopZi}
\end{align}
\end{subequations}
\noindent where $C1$ and $C3$ constrain 3D flight ranges of Alice and Bob, respectively, $C2$ and $C4$ constrain the transmit powers of Alice and Bob, respectively. Moreover, $C5$ constrains the minimum achievable rate for the MU where $R_M[n]$ needs to exceed a threshold value $R_{min}$ to ensure transmission effectiveness. In addition, $C6$ and $C7$ are the total power constraints of Alice and Bob, respectively, with $P_T$ denoting the maximum total power of UAVs over $N$ time slots. Additionally, $C8$ limits the interference temperature from Bob to the MU, with $I_{0}$ indicating the maximum interference power to ensure that the interference does not affect the communication of other maritime devices.

\subsection{Problem Analyses}

\par Furthermore, we provide the corresponding analyses of the SEMCMOP.

\par \textbf{\textit{(i) Dynamic Optimization:}} In the considered scenario, Alice dynamically adjusts its data transmissions to track the moving MU, making the data link channel time-varying. At this point, Eve continuously adjusts its position to eavesdrop on the signals from Alice. Meanwhile, Bob requires real-time adjustments based on Eve and MU for effective jamming, causing the jamming link channel to be dynamic. Thus, the SEMCMOP is a dynamic optimization problem.

\par \textbf{\textit{(ii) NP-hard Complexity:}} For simplicity in analysis, we investigate the first optimization objective under constrained operational parameters. Specifically, by fixing the positions of the MU and Eve, while quantizing the transmit powers of Alice and Bob into discrete levels, the original problem reduces to the following simplified formulation:
\begin{subequations}
\label{con:NP-prove}
\begin{align}
&\min _{\left\{ \mathbb{L}_{A},\mathbb{P}_{A},\mathbb{L}_{B},\mathbb{P}_{B} \right \} } \ F= -f_{1} , \label{con:npa}\\
\mbox{s.t.}\ & \text{Eqs.} (\ref{con:mopZb}), (\ref{con:mopZd}),  (\ref{con:mopZf})- (\ref{con:mopZi}),\\
&P_{A}[n] \in [0, P_{max}], \forall n\in \mathbb{N} , \label{con:npc}\\
&P_{B}[n] \in [0, P_{max}], \forall n\in \mathbb{N} , \label{con:npd}\\
&{\textstyle \sum_{n=1}^{N}} P_{A}[n] < N P_{max}, \forall n\in \mathbb{N} ,\label{con:npe}\\
&{\textstyle \sum_{n=1}^{N}} P_{B}[n] < N P_{max}, \forall n\in \mathbb{N} .\label{con:npf}
\end{align}
\end{subequations}

\par The reduced-form SEMCMOP constitutes a nonlinear multidimensional 0-1 knapsack configuration problem, which is explicitly categorized as NP-hard in complexity computation~\cite{Goos2020}. This complexity extends to the original SEMCMOP when the discrete constraints are generalized to continuous domains. Consequently, the SEMCMOP exhibits NP-hard complexity.

\par \textbf{\textit{(iii) Long-term Optimization Objectives:}} The continuous movements of the MU, Alice, Eve, and Bob introduce time-varying channel conditions that significantly influence the optimization objectives. Moreover, the SEMCMOP focuses on the sequential decision-making process of UAVs and aggregates objective evaluations over $N$ time slots, meaning that solutions optimized for individual time slots may perform poorly when evaluated over the complete operational duration. Consequently, the SEMCMOP features long-term optimization objectives.

\par In summary, the SEMCMOP presents unique challenges that render conventional convex optimization methods and evolutionary algorithms inadequate~\cite{Wang2023}. In this case, the DRL algorithm offers a promising alternative, as it can autonomously learn optimal policies through environmental interactions while enabling real-time decision-making~\cite{Wu2024}. Therefore, we adopt the DRL algorithm to solve the formulated SEMCMOP.

%
\section{Algorithm} 
\label{sec:algorithm}

\par In this section, we first formulate the SEMCMOP as a POMDP, followed by an introduction to the conventional SAC algorithm. Next, given the challenges of conventional SAC in POMDP, we propose an SAC-CVAE algorithm to address these challenges.

\subsection{POMDP Formulation}

\par For effective implementation of robust DRL algorithms, we transform the formulated SEMCMOP into a POMDP. Specifically, a POMDP extends the standard Markov decision process (MDP) by incorporating perceptual limitations that restrict agents from directly observing the complete state. The POMDP is structured by $\langle \mathcal{S},\mathcal{A},\mathcal{O},\mathcal{P},\mathcal{R},\gamma \rangle$~\cite{Wang2025}, where $\mathcal{S}$ denotes the global state space and $\mathcal{O}$ denotes the observation space accessible to agents. At time slot $n$, the state of the environment is represented by $ \boldsymbol{s}[n] \in \mathcal{S} $, and the observations of the agent are denoted by $ \boldsymbol{o}[n] \in \mathcal{O} $. Moreover, the action space is given by $\mathcal{A}$, including the independent action spaces of Alice and Bob. In addition, $ \mathcal{P} (\boldsymbol{s}[n + 1]|\boldsymbol{s}[n], \boldsymbol{a}[n])$ represents the probability of transitioning from state $\boldsymbol{s}[n]$ to the next state $\boldsymbol{s}[n + 1]$ after performing the action $\boldsymbol{a}[n]\in \mathcal{A}$. Then, the reward function $\mathcal{R}$ evaluates optimization objectives, and the single-slot reward at time slot $n$ is given as $r[n]$, and $\gamma \in [0,1)$ is the temporal discount factor balancing immediate versus future rewards. The action of an agent is determined by the policy $\pi$, where the probability of choosing an action in the state is expressed as $\pi(\boldsymbol{a}|\boldsymbol{s})$, and the goal of the POMDP is to determine a policy $\pi$ that maximizes cumulative rewards. Next, we introduce the necessary elements of the POMDP in detail.

\subsubsection{State}

\par In the dynamic decision-making process, the agent needs to extract the real-time state to develop a corresponding policy. This agent is concerned with multi-dimensional information, including the parameters of Alice and Bob, and the spatial coordinates of the MU and Eve. Specifically, the real-time coordinates of the vessels can be obtained by using the vessel movement model in Section \ref{sec:models_and_formulation}. However, Eve may employ adaptive techniques that prevent the agent from obtaining accurate and complete position information. To improve the jamming effectiveness of Bob, we consider predicting the unobserved positions of Eve. Note that building an effective prediction model based solely on current position samples is challenging. To this end, we introduce the historical trajectory sequence of Eve as a component of the observation space, which is expressed by
\begin{equation}
    \begin{split}
     \boldsymbol{o}[n]=\{\mathcal{L}_{E}[n-Z+1], \mathcal{L}_{E}[n-Z+2], \ldots, & \mathcal{L}_{E}[n], \\&\forall n \in \mathbb{N} \},
    \end{split}
\end{equation}
\noindent where $ \mathcal{L}_{E}[n]=\{x_{E}[n],y_{E}[n],z_{E}[n]\}$ represents the 3D location of Eve at time slot $n$, and $Z$ is the maximum storage length of the historical trajectory sequence. When $n < Z$, the length of the trajectory sequence is equal to $n$.

\par Furthermore, the global state space contains the observation space, which can be expressed by
\begin{equation}
    \begin{split}
    \mathcal{S}=\{&\boldsymbol{s}[n]|\boldsymbol{s}[n]=(\boldsymbol{\Upsilon}[n], \boldsymbol{o}[n], (x_A[n],y_A[n],z_A[n]), \\ &  (x_B[n],y_B[n],z_B[n])),P_{A}[n],P_{B}[n],\forall n \in \mathbb{N} \}.
    \end{split}
\end{equation}

\subsubsection{Action}

\par In our considered scenario, Alice and Bob need to dynamically optimize their flight trajectories and transmit powers to ensure reliable and secure maritime communications. Accordingly, the action space is denoted by
\begin{equation}
    \mathcal{A}= \{\boldsymbol{a}[n]| \boldsymbol{a}[n]= (\mathcal{A}_{A}[n],P_{A}[n],\mathcal{A}_{B}[n],P_{B}[n]),\forall n \in \mathbb{N}  \},
\end{equation}
\noindent where $\mathcal{A}_{A}[n]$ and $\mathcal{A}_{B}[n]$ are the 3D action vectors of Alice and Bob, respectively. Note that our research considers an eavesdropping UAV (Eve) and a corresponding assisted UAV (Bob). Furthermore, our approach has good scalability, enabling adaptation to extended scenarios with multiple eavesdropping UAVs and jamming UAVs.

\subsubsection{Reward}

\par The reward function serves as a critical feedback mechanism that guides agent actions and determines the quality of the policy. Thus, we design a composite reward structure comprising reward components and penalty terms. This dual mechanism ensures efficient policy exploration while maintaining operational constraints.  Note that the constraints $C1$-$C4$, $C6$, and $C7$ of the SEMCMOP are fulfilled by configuring the UAV parameters to operate within their specified allowable ranges. The remaining constraints $C5$ and $C8$ are satisfied by incorporating them into the reward function as penalty components. Specifically, we set a penalty item $W_{1}$ according to the constraint $C5$ to guarantee the transmission requirement for legitimate users as follows:
\begin{equation}
\begin{split}
W_{1}[n]& = \left\{\begin{array}{ll}
R_{M}[n], & \text{if}\hspace{0.1cm} R_{M}[n] \le R_{min}.\\
0, &\text{otherwise}.
\end{array}\right.
\end{split}
\end{equation}

\par Then, we set a penalty item $W_2$ based on the constraint $C8$ to prevent jamming signals from disrupting legitimate maritime communication as follows:
\begin{equation}
\begin{split}
&W_{2}[n] = \\&\left\{\begin{array}{ll}
P_{B}[n]G_{B}\left|\mathcal{C}_{B,M}[n]\right|^{2}, & \text{if}\hspace{0.1cm} P_{B}[n]G_{B}\left|\mathcal{C}_{B,M}[n]\right|^{2} > I_{0}.\\
0, &\text{otherwise}.
\end{array}\right.
\end{split}
\end{equation}

\par Therefore, the reward function is formulated as follows:
\begin{equation}
\begin{split}
    \mathcal{R} = \{&r[n]|r[n] = \omega_1 \mu_{1}C_{s}[n]- \omega_2\mu_{2}(E_A[n]+E_{B}[n])\\&- \mu_{3}W_{1}[n] - \mu_{4}W_{2}[n], \forall n\in \mathbb{N}  \},
\end{split}
\end{equation}
\noindent where $\mu_1$-$\mu_4$ are scaling factors to ensure that the different targets are on the same order of magnitude. Moreover, $\omega_1$ and $\omega_2$ are the weights of the two objectives, respectively.

\subsection{Conventional SAC Algorithm} 

\par Next, we discuss the advantages of the SAC algorithm in dealing with MDP and describe the process in detail.

\subsubsection{Selection of SAC Algorithm}

\par Traditional DRL algorithms, such as discrete-action approaches (\textit{e.g.}, deep Q-network (DQN)) fail to support continuous control tasks~\cite{Fan2020}. Moreover, while trust region policy optimization (TRPO) provides improved policy stability through trust region optimization, it imposes prohibitively high computational complexity~\cite{Schulman2015}. These algorithmic limitations face significant challenges when addressing the continuous and rapidly evolving POMDP.
 
\par To overcome these challenges, we adopt the SAC algorithm as our optimization framework. Specifically, its entropy maximization principle promotes systematic exploration across vast state-action spaces, avoiding premature convergence to suboptimal policies. Moreover, its twin Q-value network architecture and policy smoothing reduce value overestimation, enhancing learning stability in uncertain environments. In addition, the automated temperature adjustment mechanism dynamically balances exploration and exploitation, avoiding manual hyperparameter tuning in complex and dynamic environments. Therefore, we select SAC as the foundational framework for the POMDP.

\par The SAC algorithm introduces maximum entropy to encourage exploration, and the redefined reward function is denoted by
\begin{equation}
\begin{split}
    J(\pi) = {\sum_{n=1}^{N}}\mathbb{E} \{\gamma  \left[  r[n]  + \alpha \mathcal{H}(\pi ( \cdot |\boldsymbol{s}[n]))  \right]|\pi \},
\end{split}
\end{equation}
\noindent where $\mathbb{E}\{ \cdot \}$ is the expectation indicator, $\mathcal{H}(\pi ( \cdot |\boldsymbol{s}[n])) = -\log \pi(\boldsymbol{a}[n]|\boldsymbol{s}[n])$ is the entropy of the policy $\pi$, and $\alpha$ is the temperature parameter that controls the balance between the entropy term and reward, thereby regulating the stochasticity of the optimal policy. Furthermore, within the actor-critic architecture, the critic and actor are allocated to policy evaluation and policy optimization, respectively, as introduced below.

\subsubsection{The Critic Part}

\par The SAC algorithm effectively handles continuous action spaces by implementing an approximate form of soft policy iteration. By employing parametric approximators for both the Q-value and policy networks, this approach achieves optimization via stochastic gradient descent mechanisms. In the SAC framework, we consider three key components which consist of a state-value network $V_{\psi} (\boldsymbol{s}[n])$, a soft Q-value network $Q_{\theta} (\boldsymbol{s}[n], \boldsymbol{a}[n])$, and a tractable policy network $\pi_{\Phi} (\boldsymbol{a}[n] |\boldsymbol{s}[n])$, where $\psi, \theta,$ and $\Phi$ represent their respective network parameters.

\par To enhance training stability, a separate function approximator is set for the state-value network~\cite{Haarnoja2018}. The state-value network is trained to minimize the squared residual error as follows:
\begin{equation}
\begin{split}
    J_{V}&(\psi)=\mathbb{E}\{\frac{1}{2}[V_\psi (\boldsymbol{s}[n])-\\ &\mathbb{E}\{Q_{\theta}\left(\boldsymbol{s}[n], \boldsymbol{a}[n]\right)-\alpha \log {\pi_{\Phi}}\left(\boldsymbol{a}[n]|\boldsymbol{s}[n]\right)|\pi_{\Phi}\}]^{2} |\mathcal{D}\},
    \end{split}
\label{state-value}
\end{equation}
\noindent where $\mathcal{D}$ denotes the replay buffer, and the parameter $\psi$ undergoes iterative refinement with the stochastic gradient $\nabla_{\psi}J_{V}(\psi)$~\cite{Zhang2021}. Moreover, the soft Q-value network parameter is trained by reducing the soft Bellman residual, which is expressed by 
\begin{equation}
J_{Q}(\theta )=\mathbb{E} \{ \frac{1}{2} [ Q_{\theta} \left ( \boldsymbol{s}[n], \boldsymbol{a}[n]\right )- \hat{Q}\left ( \boldsymbol{s}[n], \boldsymbol{a}[n]\right ) ]^{2} | \mathcal{D}  \},
\label{loss-Q}
\end{equation}
\noindent where $\hat{Q}\left (\boldsymbol{s}[n], \boldsymbol{a}[n]\right ) = r[n]  + \gamma \mathbb{E}\{V_{\hat{\psi} }(\boldsymbol{s}[n+1]) \}$ is the Q target value at time slot $n$. Correspondingly, the parameter $\theta$ is optimized through stochastic gradient descent $\nabla_{\theta}J_{Q}(\theta)$.

\subsubsection{The Actor Part}

\par The primary objective of the actor component is to search for policy improvements. Our approach utilizes the state-conditional stochastic policy network $\pi$ to sample actions, and then uses the KL divergence to evaluate. Moreover, we use a neural network transformation to reparameterize the policy, resulting in a lower variance estimator. At this point, the policy network can be learned as follows:
\begin{equation}
\begin{split}
     J_{\pi}(\Phi)=\mathbb{E}\{\alpha\log \pi_{\Phi} &\left(f_{\Phi}(\epsilon[n]; \boldsymbol{s}[n]) |\boldsymbol{s}[n]\right)- \\ &Q_{\theta}\left(\boldsymbol{s}[n], f_{\Phi}(\epsilon[n] ; \boldsymbol{s}[n]) \right) | \mathcal{D}, \mathcal{N}\},
    \label{loss-pi}
\end{split}
\end{equation}
\noindent where $f_{\Phi}(\epsilon[n]; \boldsymbol{s}[n])$ is the reparameterization trick, and $\epsilon \sim \mathcal{N}(0,1)$ is an action noise signal sampled from a standard normal distribution~\cite{Zhang2021}. Similarly, the parameter $\Phi$ can be optimized with stochastic gradient $\nabla_{\Phi}J_{\pi}(\Phi)$.

\subsection{The Proposed SAC-CVAE Algorithm}

\par In this part, we present the motivation of proposing the SAC-CVAE algorithm and provide the implementation details of this algorithm.

\subsubsection{Motivation of SAC-CVAE Algorithm}

\begin{algorithm}
\label{alg:algorithm1}
\caption{SAC-CVAE Algorithm}
\KwIn {Number of iterations $I$, batch size, update rate $\tau$, and learning rates.}
\textbf{Initialize:} Replay buffer $\mathcal{D}$, critic networks $Q_\theta$ and $V_\psi$, and actor network $\pi_\Phi$\;
\For{each iteration $i = 1,2,\dots,I$}{
Initialize the environmental information\;
\For{each step $n = 1,2,\dots,N$}{
Store the observed location of Eve $\mathcal{L}_{E}[n]$\;
\If{ $n \ge Z$}{
\tcp{LSTM-assisted prediction mechanism}
LSTM processes historical trajectory sequence of Eve $\boldsymbol{o}[n]$ by Eqs.~(\ref{forget gate})-(\ref{output gate});\\
Predicts the position of Eve\;
Obtain complete observation space $\boldsymbol{o}[n]$\;
}
Obtain global state space $\boldsymbol{s}[n]$ by Eq.~(\ref{Simplified state})\;
Select and execute action $\boldsymbol{a}[n]$, $\boldsymbol{a}[n]\sim \pi_{\Phi}(\boldsymbol{a}[n]|\boldsymbol{s}[n])$\;
Update the environmental information, obtain $\boldsymbol{o}[n+1]$\;
Observe next state $\boldsymbol{s}[n+1]$ and reward $r[n]$\;
Store $\left ( \boldsymbol{s}[n], \boldsymbol{a}[n],r[n], \boldsymbol{s}[n+1] \right )$ to $\mathcal{D}$\;
Update the state-value network by Eq. (\ref{state-value})\;
Update the soft Q-value network by Eq. (\ref{loss-Q})\;
Obtain the optimized policy by \textbf{Algorithm~\ref{alg:algorithm2}};\\
Update the target network with $\hat{\psi} \leftarrow \tau \psi+(1-\tau) \hat{\psi}$\;
}
}
\KwOut{Trained model.}
\end{algorithm}

\par While the SAC algorithm can solve continuous-time problems, it faces the following challenges when dealing with the POMDP.

\par \textit{(i) Suboptimal Solutions in the Multi-modal Decision Space:} In our considered dynamic scenario, a single state may correspond to multiple distinct yet optimal actions (\textit{i.e.}, a multi-modal decision space), where each action may lead to different future states and rewards. While the conventional SAC algorithm encourages exploration through entropy regularization, it fails to explicitly distinguish or model different action modalities~\cite{Haarnoja2018}. This limitation causes its learned policy to converge toward a broad peak distribution that inappropriately averages across potentially optimal actions, exhibiting suboptimal or unstable solutions. Note that this action-averaging issue becomes particularly detrimental in long-term trajectory optimization tasks, where decisively selecting one specific action modality rather than blending viable alternatives is crucial for achieving stable and optimal performance trajectories.

\par \textit{(ii) Computational inefficiency in the High-dimensional State Space:} The optimization problems under consideration involve high-dimensional state spaces. In particular, the historical trajectory sequence of Eve is utilized to predict its unobserved position, forming the observation space and being stored in the global state space. However, this significantly expands the state space dimensions, thereby increasing the computational overhead for policy updates. Notably, this challenge becomes more severe in real-time deployment systems, as excessive state dimensions can lead to latency, instability, and convergence failure.

\par
\begin{figure*}[!t]
\centering
\includegraphics[width=7in]{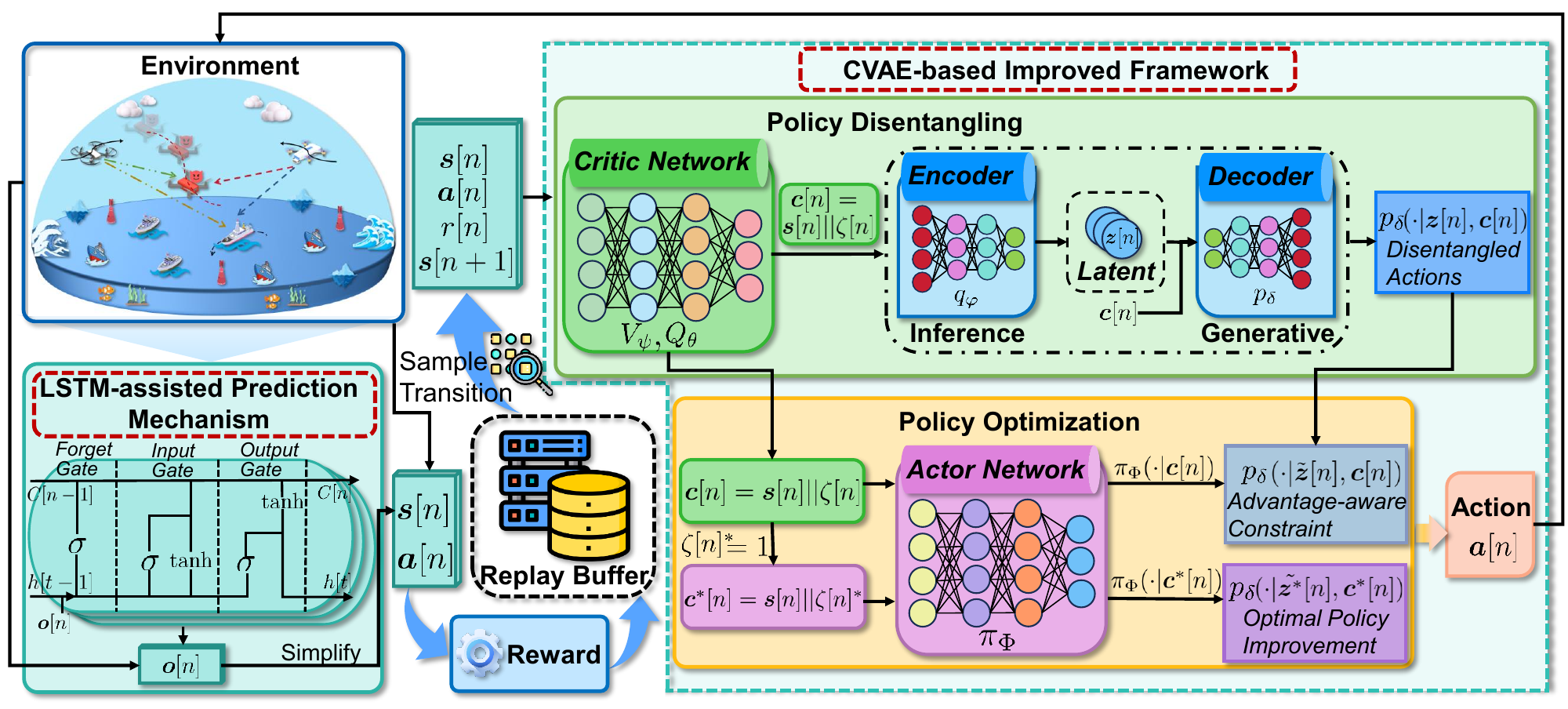}
\caption{The architecture of the proposed SAC-CVAE algorithm for solving the SEMCMOP, which integrates a CVAE-based improved framework to disentangle and optimize policies as well as an LSTM-assisted prediction mechanism to enhance computational efficiency.}
\label{fig:outline}
\end{figure*}

 To address these challenges of the conventional SAC algorithm in POMDP, we propose a novel improved algorithm, SAC-CVAE. Note that the SAC-CVAE algorithm is trained on servers and deployed to UAVs for execution, thereby balancing computational demands with constrained UAV resources. Furthermore, Fig. \ref{fig:outline} provides the visual architecture of the SAC-CVAE algorithm, and Algorithm \ref{alg:algorithm1} outlines its overall structure. The main advances in the SAC-CVAE algorithm are detailed in the following sections.

\subsubsection{Conditional Variational Autoencoder (CVAE)-based Improved Framework}

\par Variational autoencoder (VAE) provides a principled framework for learning latent representations of data~\cite{Sun2025}. By combining an encoder-decoder architecture with variational inference, VAE can generate diverse samples while maintaining a meaningful structure in the latent space. On this basis, we propose a CVAE-based improved framework, which disentangles policies and further optimizes the advantage-aware policies toward high advantage values, thereby avoiding the policy becoming overly biased toward a single mode. Algorithm \ref{alg:algorithm2} presents the comprehensive CVAE-based improved framework, with the implementation details elaborated below.

\par First, our framework models the advantage value as a conditional variable. Specifically, the encoder $q_{\varphi}(\boldsymbol{z}|\boldsymbol{a},\boldsymbol{c})$ processes condition $\boldsymbol{c}$ and action $\boldsymbol{a}$ to generate a latent representation $\boldsymbol{z}$, and the corresponding decoder $p_{\delta}(\boldsymbol{a}|\boldsymbol{z},\boldsymbol{c})$ reconstructs $\boldsymbol{a}$ by preserving the correlation between $\boldsymbol{c}$ and $\boldsymbol{z}$. Notably, unlike previous state-conditioned methods~\cite{Chen2022}, our framework incorporates the state $\boldsymbol{s}$ and advantage values $\zeta$, forming a dual-conditioned input structure to enhance decision context. The state-advantage condition is given by~\cite{Qing2024}
\begin{equation}
    \boldsymbol{c}[n] = \boldsymbol{s}[n] || \zeta[n],
\label{c}
\end{equation}
\noindent where $||$ denotes vector concatenation, and the advantage value $\zeta$ can be computed by
\begin{equation}
    \zeta[n] = \tanh(Q_{\theta}(\boldsymbol{s}[n],\boldsymbol{a}[n])-V_{\psi}(\boldsymbol{s}[n])),
\label{eth}
\end{equation}
\noindent where $\tanh (\cdot)$ function is used to normalize the advantage condition.

\par Furthermore, the CVAE is trained by maximizing the evidence lower bound (ELBO) for minibatches of the state-advantage condition $\boldsymbol{c}$ and the corresponding action $\boldsymbol{a}$. The training function is defined by~\cite{Sohn2015}
\begin{equation}
\begin{split}
    J_{C}(\varphi, \delta)=&-\mathbb{E}\{\mathbb{E}\left[\log \left(p_{\delta}(\boldsymbol{a}[n] \mid \boldsymbol{z}[n], \boldsymbol{c}[n])\right)|q_{\varphi}\right]+\\&\imath  \cdot KL\left[q_{\varphi}(\boldsymbol{z}[n] \mid \boldsymbol{a}[n], \boldsymbol{c}[n]) \| p(\boldsymbol{z}[n])\right]\mathcal{D}\},
\label{cvae}
\end{split}
\end{equation}
\noindent where $\imath$ is the coefficient for balancing the KL-divergence loss term, and $p(\boldsymbol{z})$ $\sim \mathcal{N}(0, 1)$ is the latent prior. The first reconstruction term ensures accurate action generation, while the KL divergence term makes the latent representation distribution match the prior distribution.

\par During each training iteration, the critic networks $Q_{\theta}$ and $V_{\psi}$ evaluate state-action pairs to compute corresponding advantage values $\zeta$ via Eq.~(\ref{eth}). Then, the advantage-aware CVAE is trained according to Eq. (\ref{cvae}). In this case, the latent representation $\boldsymbol{z}$ captures the underlying structure of action distributions, while the state-advantage condition $\boldsymbol{c}$ guides the model to generate actions that are positively correlated with $\zeta$. Consequently, the trained CVAE can generate disentangled actions $\boldsymbol{a} \sim p_{\delta}(\boldsymbol{a}|\boldsymbol{z}, \boldsymbol{c})$ and capture the correlations between the action distribution and $\zeta$. Subsequently, the trained CVAE is used for policy optimization, generating progressively higher-quality actions.

\begin{algorithm}
\label{alg:algorithm2}
\caption{CVAE-based Improved Framework}
\KwIn {CVAE training step $K$.}
\textbf{Initialize:} CVAE encoder $q_{\varphi}$ and decoder $p_{\delta}$\;
Calculate the advantage value $\zeta[n]$ by Eq. (\ref{eth});\\
Calculate the state-advantage condition $\boldsymbol{c}[n]$ by Eq. (\ref{c});\\
\tcp{Policy disentangling}
\If{ $i \le K$}{
Sample the latent representation $\boldsymbol{z}[n]$\;
Optimize CVAE encoder $q_{\varphi}$ and decoder $p_{\delta}$ according to Eq. (\ref{cvae}); 
}
\tcp{Policy optimization}
Optimize critic networks $Q_\theta$ and $V_{\psi}$ by optimal action $\boldsymbol{a}^*_\zeta$;\\
Optimize the advantage-aware policy toward high advantage values by Eq.~(\ref{loss-pi refine}). \\
\KwOut{Trained policy.}
\end{algorithm}

\par Then, during the policy optimization phase, we employ a hierarchical constraint to enable advantage-aware exploration. Specifically, the actor network $\pi_{\Phi}$ generates a latent representation $\tilde{\boldsymbol{z}}$ based on the condition $\boldsymbol{c}$. Then, this representation $\tilde{\boldsymbol{z}}$ is decoded into an action that aligns with its advantage values $\zeta$. These processes are denoted by
\begin{subequations}
\begin{align}
    &\tilde{\boldsymbol{z}}[n] \sim \pi_{\Phi}(\cdot \mid \boldsymbol{c}[n]),\\
    &\boldsymbol{a}_{\zeta}[n] \sim p_{\delta}(\cdot \mid \tilde{\boldsymbol{z}}[n], \boldsymbol{c}[n]).
\end{align}
\end{subequations}

\par In this case, $\pi_{\Phi}$ generates actions of different qualities that are correlated with a specified $\zeta$. Among them, the optimal action $\boldsymbol{a}^*_\zeta$ is obtained by processing the condition $\boldsymbol{c}^* = \boldsymbol{s} \| \zeta^*$, where $\zeta^* = 1$ represents the maximum advantage value. This approach optimizes the advantage-aware policy toward high advantage values. The policy network can be updated by
\begin{equation}
\begin{split}
    J_{\pi}(\Phi)= \mathbb{E}\{&-\lambda Q_{\theta}(\boldsymbol{s}[n], \boldsymbol{a}_{\zeta}^{*}[n])+\left(\boldsymbol{a}[n]-\boldsymbol{a}_{\zeta}[n]\right)^{2}+\\& \alpha\log \pi_{\Phi}\left(\boldsymbol{a}_{\zeta}^*[n] \mid \boldsymbol{c}[n]\right)|\mathcal{D},p_{\delta}\},
    \label{loss-pi refine}
\end{split}
\end{equation}
\noindent where $\lambda$ is the normalization coefficient to maintain proper scaling between the Q-value maximization and policy regularization. Moreover, the first term drives optimal actions through the fixed high-advantage condition $\boldsymbol{c}^*$, the second term imposes constraints on the advantage-aware policy to ensure that selected actions follow the advantage condition, and the third term is the maximum entropy term based on $\boldsymbol{c}^*$. Thus, suboptimal samples with a low advantage value $\zeta$ do not undermine the optimization of the optimal policy $\pi_{\Phi}(\cdot|\boldsymbol{c}^*)$. Instead, they impose effective constraints on the corresponding policy $\pi_{\Phi}(\cdot|\boldsymbol{c})$. This hierarchical constraint enables stable and efficient learning, where lower-quality samples guide exploration, while higher-quality actions refine policies toward optimal performance.

\par In summary, the CVAE-based improved framework combines policy disentanglement with advantage-aware policy optimization. This framework captures multi-modal action distributions and further optimizes policies toward high advantage values, improving the robustness and efficiency of our algorithm.

\subsubsection{LSTM-assisted Prediction Mechanism}

\par The historical trajectory sequence of Eve used for prediction, as an observation space, is stored in the state space, which imposes a significant computational burden. To address this challenge, we propose an LSTM-assisted prediction mechanism that calculates predictions in advance and simplifies the stored observation space to the current position of Eve. Specifically, the LSTM network is a specialized variant of recurrent neural networks and can efficiently capture temporal dependencies through its gate mechanisms~\cite{Sagheer2019}. This architecture enables it to identify complex patterns in trajectory sequences, including acceleration patterns, directional tendencies, and recurring motion sequences. Moreover, the LSTM network can selectively retain important historical information and filter irrelevant noise, thereby making it suited for modeling trajectory sequences over time. As such, we utilize the LSTM network to extract patterns from the historical trajectory sequence of Eve and predict its unobserved positions. The LSTM network architecture, as illustrated in the lower left segment of Fig.~\ref{fig:outline}, consists of three principal gates, each performing distinct functions as follows:

\par \textit{(i) Forget gate:} The forget gate ($LSTM_f[n]$) determines the amount of previous information to be discarded, which is denoted by
\begin{equation}
LSTM_f[n] = \sigma(W_f \cdot [h[n-1], \boldsymbol{o}[n]] + b_f),
\label{forget gate}
\end{equation}
\leavevmode\noindent where $\sigma(\cdot)$ is the sigmoid function to control output values in $[0,1]$, with $0$ indicating complete discarding and $1$ representing full preservation of the previous cell state $C[n-1]$. Moreover, $W_f$ and $b_f$ are the weight matrix and bias vector for the forget gate, respectively, and $h[n-1]$ is the hidden state of the previous time slot.

\par \textit{(ii) Input gate:} The input gate ($LSTM_i[n]$) regulates updates to the cell state through the following two operations:
\begin{subequations}
\label{input gate}
\begin{align}
&LSTM_i[n] = \sigma(W_i \cdot [h[n-1], \boldsymbol{o}[n]] + b_i), \\
&\tilde{C}[n] = \tanh(W_C \cdot [h[n-1], \boldsymbol{o}[n]] + b_C),
\end{align}
\end{subequations}
\noindent where $\tilde{C}[n]$ denotes the new candidate values for state updates. Moreover, $W_i$ and $W_C$ are the weight matrices for input components, $b_i$ and $b_C$ are the corresponding bias vectors. Then, the cell state updates via $C[n] = LSTM_f[n] \cdot C[n-1] + LSTM_i[n] \cdot \tilde{C}[n]$.

\par \textit{(iii) Output gate}: The output gate ($LSTM_o[n]$) generates output information as follows:
\begin{subequations}
\label{output gate}
\begin{align}
& LSTM_o[n] = \sigma(W_o \cdot [h[n-1], \boldsymbol{o}[n]] + b_o), \\
& h[n] = LSTM_o[n] \cdot \tanh(C[n]),
\end{align}
\end{subequations}
\noindent where $h[n]$ becomes the final hidden state containing distilled sequential information, which is then fed into a fully connected layer to generate the predicted position.

\par At this point, we can obtain the prediction results using the LSTM network, and the simplified global state space is denoted by
\begin{equation}
    \begin{split}
    \mathcal{S}=\{&\boldsymbol{s}[n]|\boldsymbol{s}[n]=(\boldsymbol{\Upsilon}[n],\mathcal{L}_{E}[n],(x_A[n],y_A[n],z_A[n]), \\ &  (x_B[n],y_B[n],z_B[n])),P_{A}[n],P_{B}[n],\forall n \in \mathbb{N} \},
    \end{split}
    \label{Simplified state}
\end{equation}
\noindent where $\mathcal{L}_{E}$ denotes either the observed location of Eve when available or the predicted location of Eve obtained by the LSTM network based on the historical trajectory sequence of Eve. Note that the LSTM network can be periodically fine-tuned with newly collected historical data to adapt to evolving movement patterns.

\par In summary, we employ the LSTM network that selectively filters information via its gating mechanisms while preserving relevant historical features in memory cells, to predict the unobserved position of Eve. Furthermore, we compress the state space from the historical trajectory sequence of Eve to the current predicted (or observed) position of Eve, significantly reducing the state space dimension and improving the computational efficiency of our algorithm.

\subsection{Complexity Analysis of SAC-CVAE Algorithm}

\par In this part, we provide a comprehensive analysis of the resource requirements of the SAC-CVAE algorithm, including computational complexity and space complexity. 

\par The computational complexity of the SAC-CVAE algorithm can be decomposed into the following four major components.

\par \textit{(i) Network Initialization:} The network setup requires parameter initialization. The corresponding complexity is $ \mathcal{O}(2|\theta|+|\psi|+|\Phi|)$, where $|\theta|$ and $|\psi|$ represent the number of parameters in each of the twin Q-value networks and state-value network, respectively, and $|\Phi|$ denotes the number of actor network parameters.

\par \textit{(ii) Policy Execution:} Action selection through the policy network has the complexity of $\mathcal{O} (IN (|\Phi| + |d_{h}|^2+|\boldsymbol{z}|))$, where $I$ is the total training iterations, $N$ denotes the number of steps per iteration, $|d_{h}|$ is the dimension of the LSTM hidden state, and $|\boldsymbol{z}|$ denotes the dimension of the latent representation in CVAE.

\par \textit{(iii) Replay Buffer Collection:} The complexity of collecting transitions in the replay buffer is $\mathcal{O}(INB)$, where $B$ is the environmental interaction complexity.

\par \textit{(iv) Network Update:} For critic and actor network updates, including the advantage-aware policy network optimization, the complexity is $\mathcal{O} (2IG(2|\theta|+|\psi|+|\Phi|))$, where $G$ denotes the gradient steps per update.

\par Combining these components, the aggregate computational complexity is $ \mathcal{O}(2|\theta|+|\psi|+|\Phi|+IN (|\Phi| + |d_{h}|^2+|\boldsymbol{z}|)+INB+2IG (2|\theta|+|\psi|+|\Phi|))$.

\par The space complexity of the SAC-CVAE algorithm primarily consists of network parameters and replay buffer storage. For network architecture, the complexity is $\mathcal{O}(2|\theta|+|\psi|+|\Phi|+|\boldsymbol{z}|+|d_{h}|^2)$ for critic and actor networks, latent representation, along with the LSTM hidden state. Moreover, the replay buffer stores current states, actions, rewards, and next states. Given a replay buffer capacity $D$, the complexity is $D(2|\boldsymbol{s}|+|\boldsymbol{a}|+1)$, where $|\boldsymbol{s}|$ and $|\boldsymbol{a}|$ denote the state dimension and action dimension, respectively. Thus, the aggregate space complexity is $\mathcal{O}(2|\theta|+|\psi|+|\Phi|+|\boldsymbol{z}|+ |d_{h}|^2 +D(2|\boldsymbol{s}|+|\boldsymbol{a}|+1))$.

%
\section{Simulation results and analyses}
\label{sec:simulation-results-and-analysis}

\par In this section, we evaluate the performance of the SAC-CVAE algorithm through simulation results.

\subsection{Simulation Configurations}

\par In this part, we detail the parameter configurations adopted for our simulations and present the baselines selected for comparative evaluation.

\subsubsection{Parameter Configurations}

\par We execute all simulations on a high-performance computing platform equipped with an AMD EPYC 7642 48-Core processor, NVIDIA GeForce RTX 3090 graphics card, and 128GB system memory. 

\par In the simulations, the development environment is Python 3.8 and Visual Studio Code 1.91. The UAVs (Alice and Bob) are initialized within a 100 m $\times$ 100 m area, with randomized starting positions to simulate real conditions where UAVs might be transitioning from previous tasks. Meanwhile, for the SAC-CVAE algorithm, each actor and critic network has two hidden-layer architectures with the ReLU activation function, and the Adam optimizer for parameter updates. Moreover, the batch size is 128 from the replay buffer, and the remaining main parameters are shown in Table \ref{table：table1}.

\begin{table}[htbp]
\caption{Main parameters in the simulation process}
\label{table：table1}
\begin{center} 
\renewcommand\arraystretch{1.1}
\newcommand{\tabincell}[2]{\begin{tabular}{@{}#1@{}}#2\end{tabular}}
\begin{tabular}{lll}
\noalign{\hrule height 1pt}
\textbf{Notation} & \textbf{Definition} & \textbf{Value} \\
\hline 
$d_{r}$ & Reference distance of the U2V link & 2600 m\\
$P_T$ & Maximum total power of the UAV & 400 mW\\
$F_{V}$ & Rician factor & 31.3 \\
$f_{c}$ & Carrier frequency & 2.4 GHz \\
$\gamma$ & Discount factor & 0.9 \\
$G_{A}$ & Antenna gain of Alice & 8 dBi \\
$G_{B}$ & Antenna gain of Bob & 8 dBi \\ 
$I_{0}$ & Maximum interference power & -74 dBm\\
$I_{r}$ & Path loss relevant index & 1.5\\
${I}_{h}$ & MU horizontal inertial matrix element & 300 kg$\cdot m^2$ \\
${I}_{z}$ & MU vertical inertial matrix element & 150 kg$\cdot m^2$ \\
$m_{MU}$ & Mass of the MU & 100 kg \\
$m_{U}$ & Mass of the UAV & 2 kg\\
$P_{d}$ & Path loss parameter of the U2V link & 116.7 \\ 
$R_{min}$ & Threshold value of effective transmission & 0.0014 \\
$\sigma^{2}$ & Power of additive white Gaussian noise & -107 dBm\\
$v_{l}$ & Linear velocity of the MU & 1 m/s \\
$v_r$ & Rotational velocity of the MU & 0.5 rad/s\\
$x_A^{min}$ & Minimum x-coordinate of Alice & 100 m\\
$x_A^{max}$ & Maximum x-coordinate of Alice & 200 m\\
$x_B^{min}$ & Minimum x-coordinate of Bob & 200 m\\
$x_B^{max}$ & Maximum x-coordinate of Bob & 300 m\\
$y_A^{min}$ & Minimum y-coordinate of Alice & 100 m\\
$y_A^{max}$ & Maximum y-coordinate of Alice & 200 m\\
$y_B^{min}$ & Minimum y-coordinate of Bob & 400 m\\
$y_B^{max}$ & Maximum y-coordinate of Bob & 500 m\\
$z^{min}$ & Minimum altitude of Alice and Bob & 50 m\\
$z^{max}$ & Maximum altitude of Alice and Bob & 70 m\\
\noalign{\hrule height 1pt}
\end{tabular}
\end{center}
\end{table}

\subsubsection{Baselines}

\par To comprehensively assess the effectiveness of the SAC-CVAE algorithm, we provide a comparative approach and several comparison algorithms as follows:

\par \textit{(i) Non-jamming Approach:} In this scenario, Alice sends signals to the MU without jamming. This approach highlights the necessity of UAV-assisted intelligent jamming against eavesdroppers in low-altitude maritime communications.

\par \textit{(ii) State-of-the-Art DRL Algorithms:} To further evaluate the performance of SAC-CVAE, we choose conventional SAC and the following state-of-the-art algorithms as benchmarks. Specifically, the deep deterministic policy gradient (DDPG) combines policy gradient methods with deep learning, utilizing an actor-critic framework to enhance policy learning~\cite{Qiu2019}. Twin delayed DDPG (TD3) is an enhanced variant of DDPG that improves stability via double Q-learning, delayed policy updates, and target policy smoothing~\cite{Fujimoto2018}. Moreover, proximal policy optimization (PPO) optimizes policy updates with a clipping mechanism to ensure training stability and efficiency~\cite{Schulman2017}. In addition, the greedy algorithm makes locally optimal decisions at each step by maximizing immediate rewards~\cite{Shafique2005}. Additionally, all algorithms are trained for $4 \times 10^5$ training iterations, with performance evaluations conducted every 80 iterations.

\subsection{Simulation Results}

\par In this part, we evaluate the performance of the SAC-CVAE algorithm. We consider two distinct eavesdropper movement patterns, including the case where Eve approaches the MU to enhance eavesdropping capabilities and the case where Eve moves away from the MU to escape detection for future eavesdropping. For each pattern, we provide detailed analyses of jamming effectiveness, optimized objective values, convergence performance, and trajectory results. Additionally, we present supplementary performance results of our proposed approach in an extended scenario.

\subsubsection{Comparisons with Non-jamming Approach} 

\begin{figure}
\centering
\includegraphics[width=3.5in]{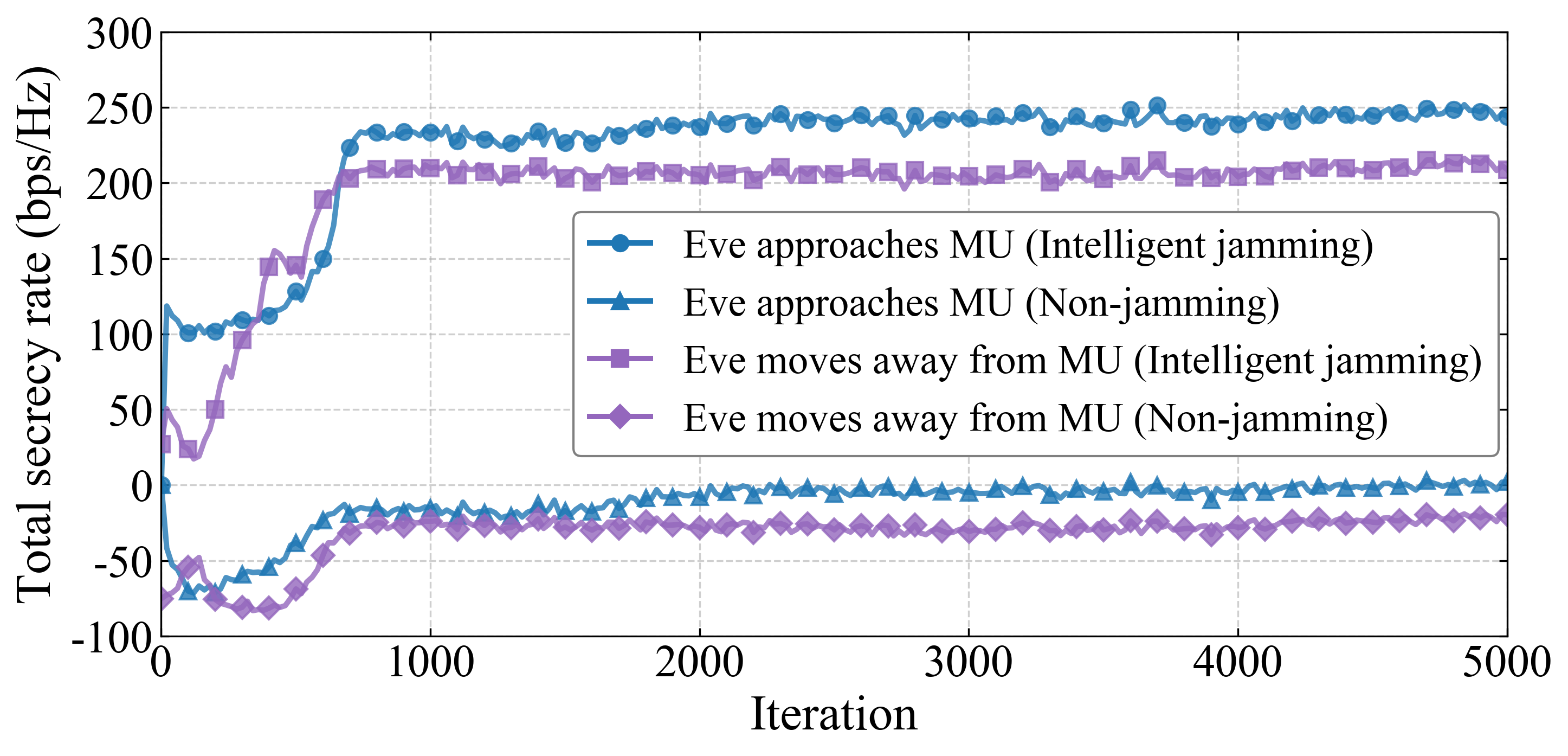}
\caption{Total secrecy rates obtained by the intelligent jamming and non-jamming approaches as Eve approaches the MU and Eve moves away from the MU.}
\label{fig:nojamming_close}
\end{figure}

\par We compare the security performance of UAV-assisted intelligent jamming and non-jamming approaches in low-altitude maritime communications. Specifically, Fig. \ref{fig:nojamming_close} presents the total secrecy rate obtained for both approaches as Eve approaches the MU and Eve moves away from the MU, where the secrecy rate at time slot $n$ in the non-jamming approach is given by $C_{s}[n]=R_{M}[n]-R_{E}[n]$ for comparative analysis. The results demonstrate that the intelligent jamming approach sustains high secrecy rates and ensures communication reliability, whereas the results of the non-jamming method are around 0. These comparative results validate the effectiveness of the UAV-assisted intelligent jamming approach in achieving secure low-altitude maritime communications.

\begin{figure}[ht]
\centering
\includegraphics[width=3.5in]{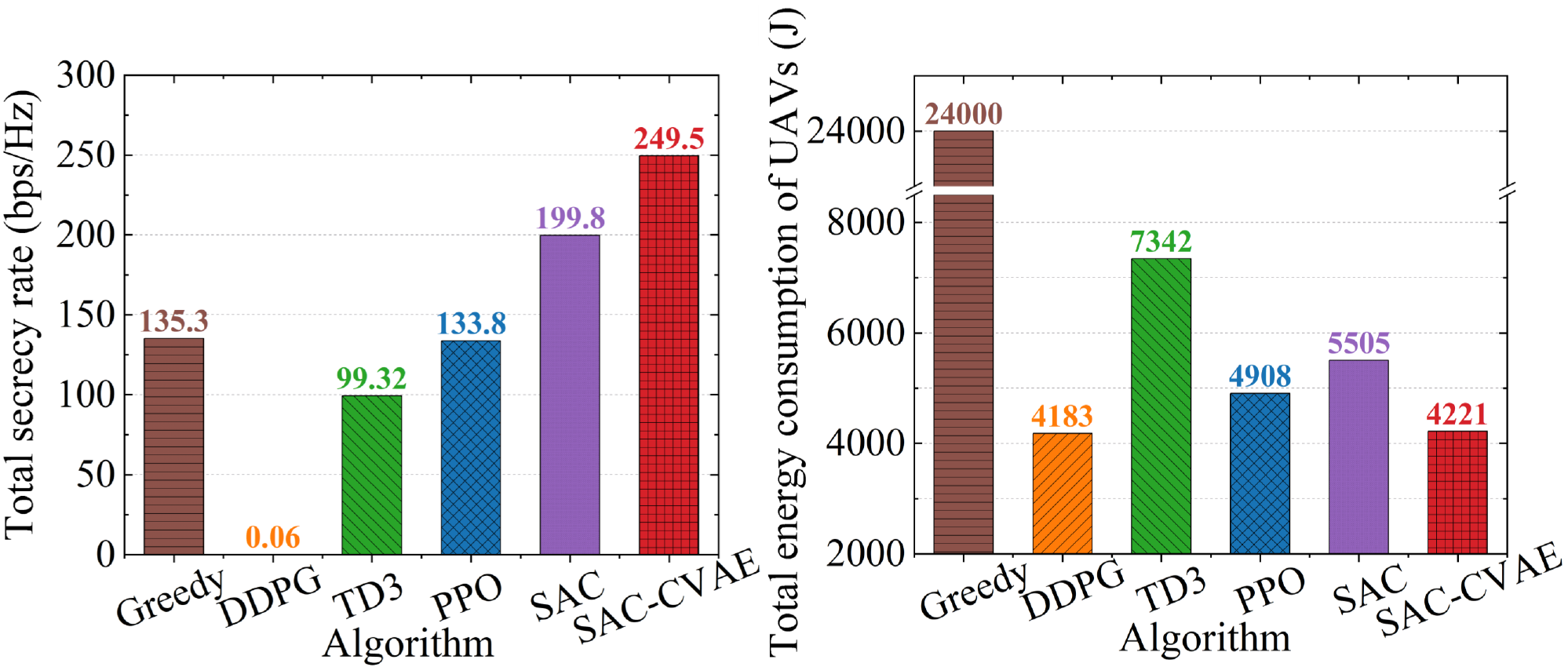}
\caption{The optimization objective values obtained by different algorithms as Eve approaches the MU.}
\label{fig:obj_close}
\end{figure}

\begin{figure}[ht]
\centering
\includegraphics[width=3.5in]{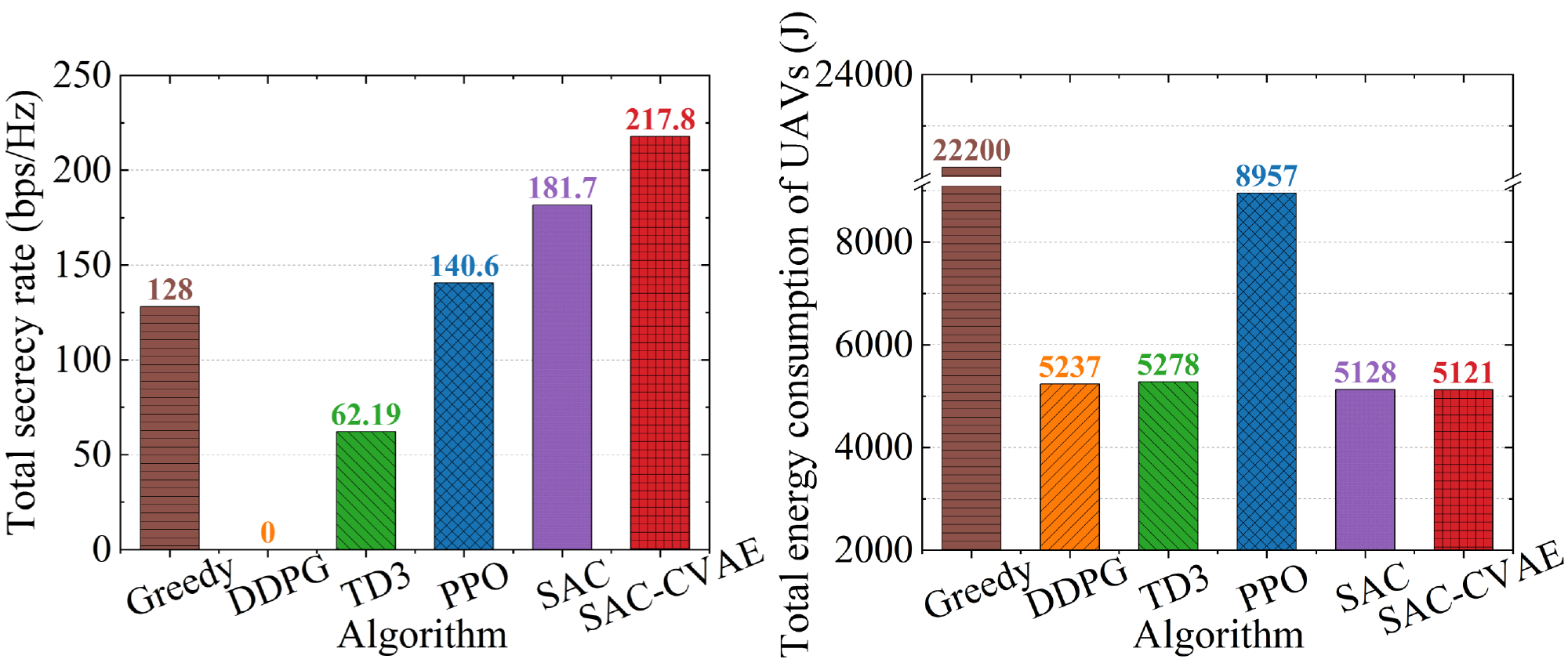}
\caption{The optimization objective values obtained by different algorithms as Eve moves away from the MU.}
\label{fig:obj_far}
\end{figure}

\subsubsection{Comparisons with Other Algorithms}

\par We evaluate the optimized objective values obtained by different algorithms. As shown in Fig. \ref{fig:obj_close}, when Eve approaches the MU, the SAC-CVAE algorithm achieves a maximum total secrecy rate and a near-optimal total energy consumption of the UAVs. While DDPG has a minimum total energy consumption, it fails to provide an effective secrecy rate. Thus, the proposed SAC-CVAE algorithm demonstrates superior performance. In addition, Fig. \ref{fig:obj_far} provides the optimized objective values as Eve moves away from the MU. The SAC-CVAE consistently outperforms other algorithms both in secrecy rate and energy consumption. This balance makes SAC-CVAE particularly suitable for secure maritime communications under the energy constraints of UAVs. These results further demonstrate that the SAC-CVAE algorithm can effectively maintain communication security while addressing the deployment challenges of UAVs in maritime environments.

\subsubsection{Convergence Performance}

\par
\begin{figure}[b]
\centering
\includegraphics[width=3.5in]{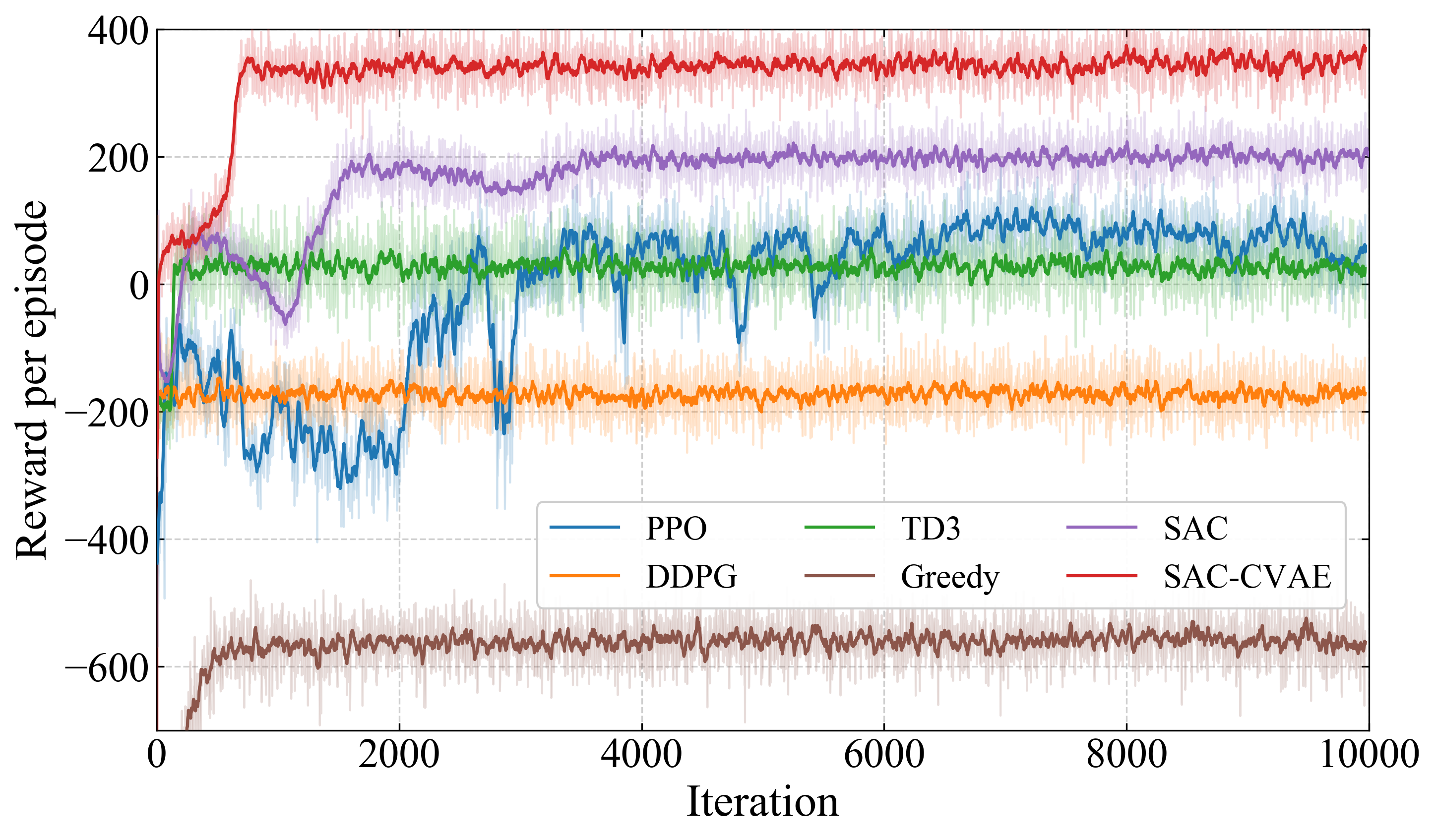}
\caption{Convergence performance obtained by different algorithms as Eve approaches the MU.}
\label{fig:reward_close}
\end{figure}

\begin{figure}
\centering
\includegraphics[width=3.5in]{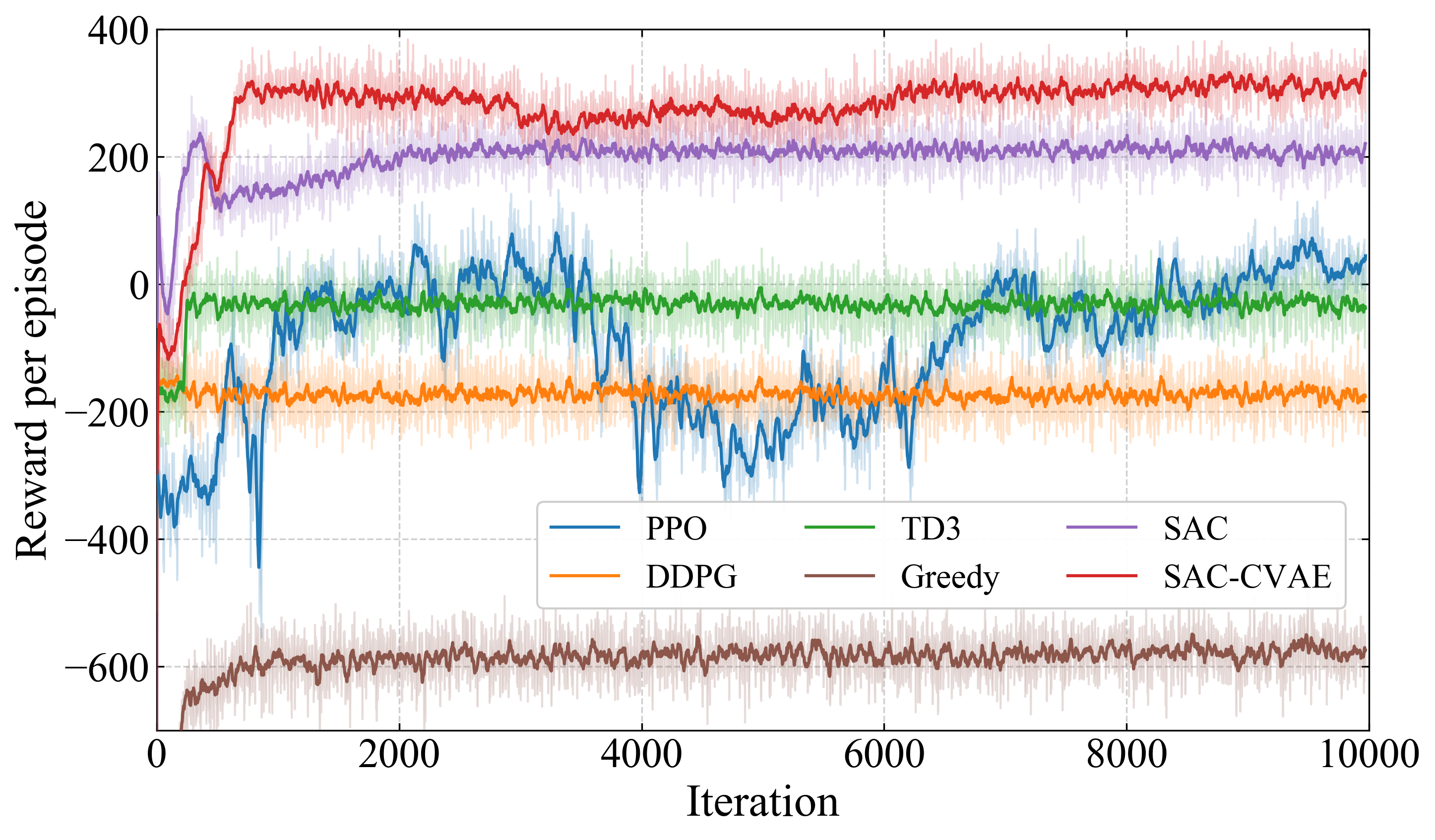}
\caption{Convergence performance obtained by different algorithms as Eve moves away from the MU.}
\label{fig:reward_far}
\end{figure}

 Convergence performance is a key metric for assessing the stability and optimization capability of DRL algorithms. Accordingly, we present convergence results of different algorithms to provide a comparative analysis. As illustrated in Fig. \ref{fig:reward_close}, when Eve approaches the MU, upon convergence, the SAC-CVAE algorithm achieves greater cumulative rewards compared to other comparison algorithms, demonstrating its superior learning efficiency. Moreover, when Eve moves away from the MU, as shown in Fig. \ref{fig:reward_far}, the converged SAC-CVAE algorithm maintains the optimal performance in terms of reward values. Therefore, the excellent convergence performance across various eavesdropper movement patterns confirms the robustness of the SAC-CVAE algorithm, further validating its ability to learn more effective policies.

\subsubsection{Trajectory Results} 

\par
\begin{figure}
\centering
\includegraphics[width=3.2in]{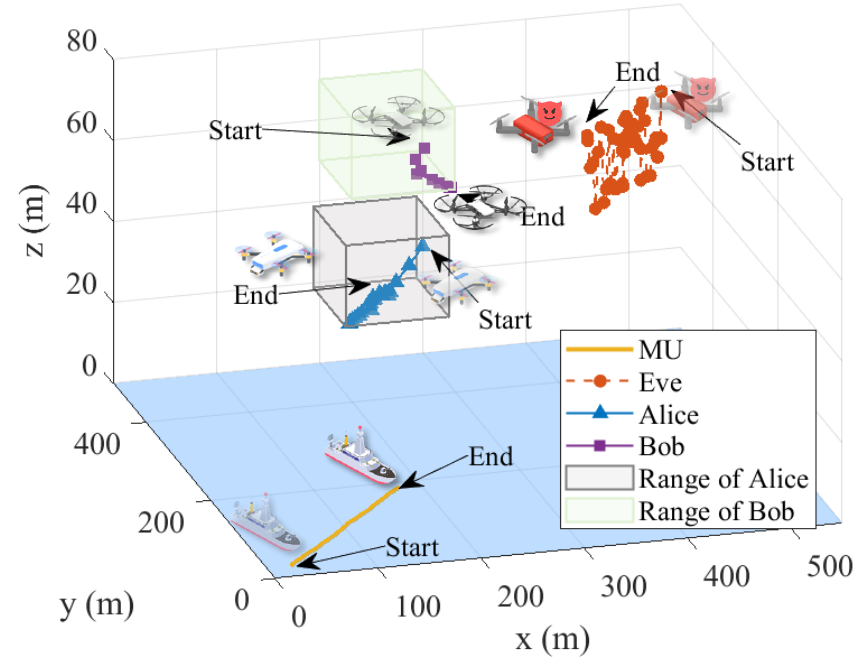}
\caption{Trajectory results obtained by the SAC-CVAE algorithm as Eve approaches the MU.}
\label{fig:trajectory_close}
\end{figure}

\begin{figure}[t]
\centering
\includegraphics[width=3.2in]{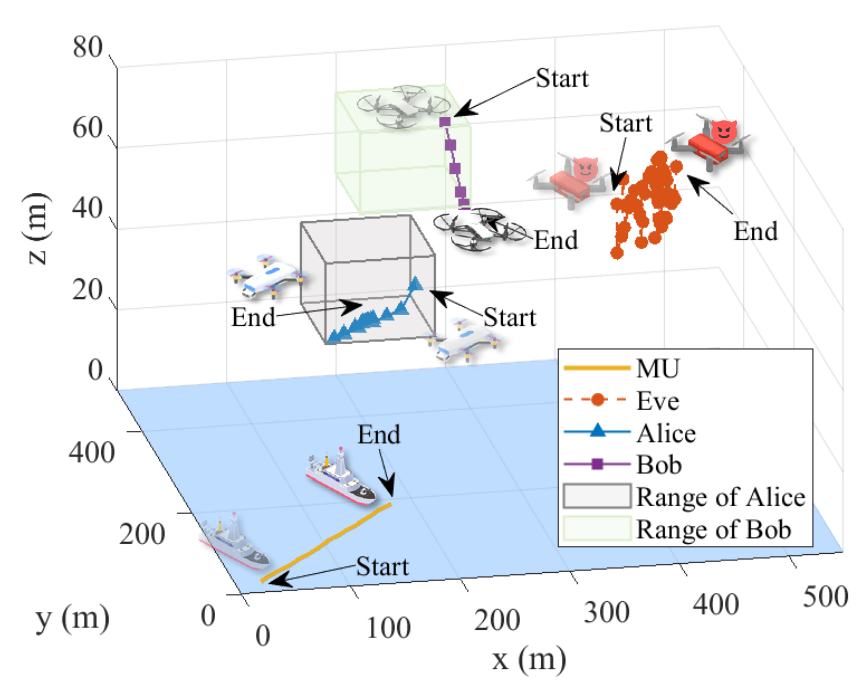}
\caption{Trajectory results obtained by the SAC-CVAE algorithm as Eve moves away from the MU.}
\label{fig:trajectory_far}
\end{figure}

 Furthermore, Figs. \ref{fig:trajectory_close} and \ref{fig:trajectory_far} illustrate the 3D trajectory results obtained by the SAC-CVAE algorithm across various eavesdropper movement patterns. Specifically, when Eve approaches the MU (Fig. \ref{fig:trajectory_close}) or moves away from the MU (Fig. \ref{fig:trajectory_far}), Alice advances toward the MU to optimize data transmission, while Bob dynamically positions toward Eve to improve jamming effectiveness. These coordinated movements indicate that Alice and Bob can adaptively track their respective targets and execute autonomous path optimization. Note that the system incorporates a security constraint preventing Eve from getting too close to Alice. Thus, the trajectory results demonstrate that the proposed SAC-CVAE algorithm can achieve intelligent trajectory optimization to enable secure low-altitude maritime communications.

\subsubsection{Extended Scenario with Multiple Eavesdropping UAVs and Jamming UAVs}

\par
\begin{figure}
\centering
\includegraphics[width=3.5in]{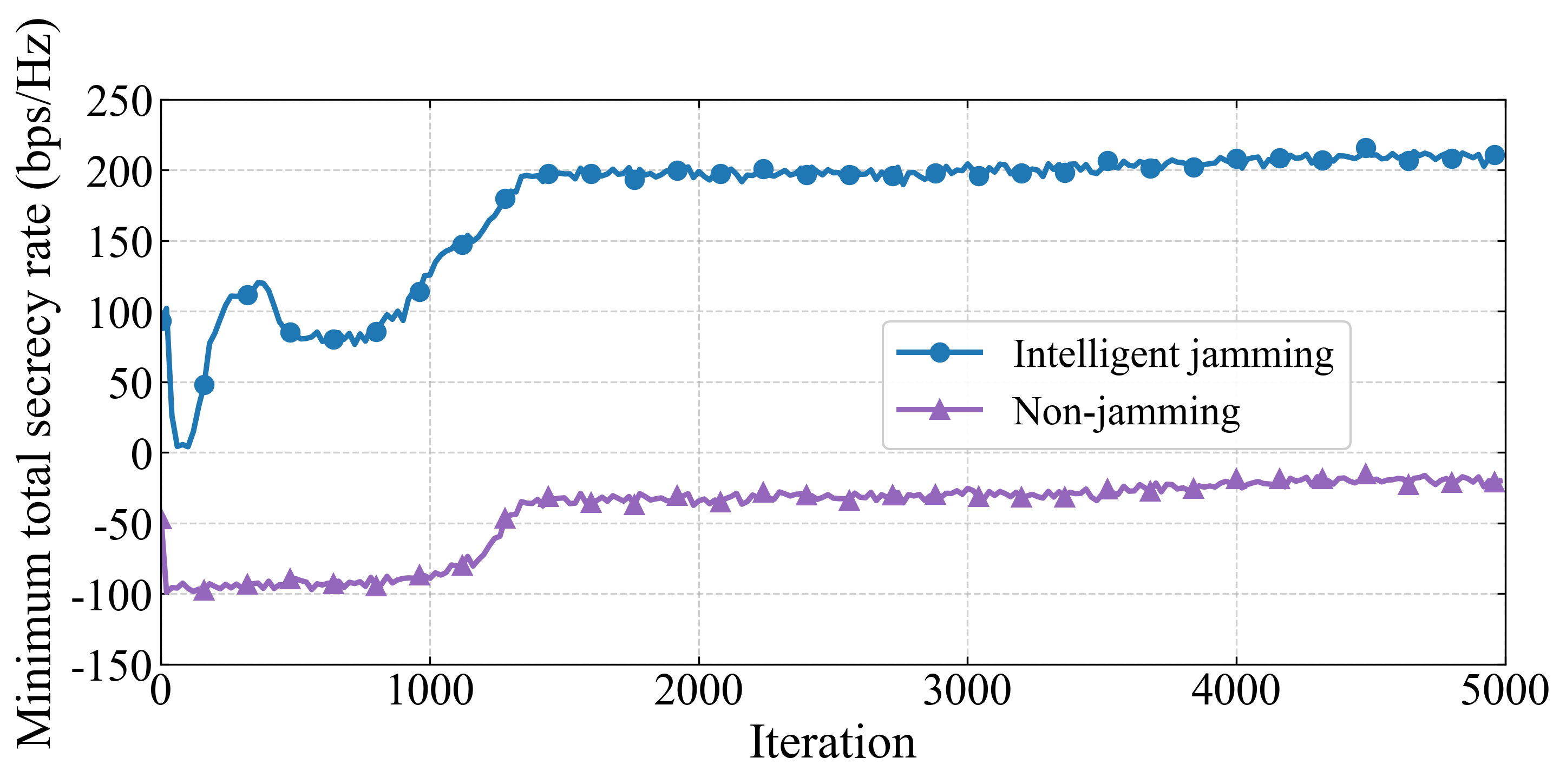}
\caption{Minimum total secrecy rates obtained by the intelligent jamming and non-jamming approaches in the extended scenario.}
\label{fig:non_jamming_multi_UAV}
\end{figure}

 Furthermore, we evaluate our approach in an extended scenario involving multiple jamming UAVs sending jamming signals to multiple eavesdropping UAVs, respectively. At this point, the first optimization problem in the SEMCMOP becomes maximizing the minimum total secrecy rate of the system, and the POMDP needs to incorporate expanded state and action spaces to accommodate multiple eavesdropping UAVs and jamming UAVs. Moreover, Fig.~\ref{fig:non_jamming_multi_UAV} presents the minimum total secrecy rates of the system obtained by the intelligent jamming and non-jamming approaches in the extended scenario. As can be seen, our proposed intelligent jamming approach obtains a superior minimum total secrecy rate, thereby ensuring reliable and secure low-altitude maritime communications. In contrast, the minimum total secrecy rate obtained by the non-jamming approach is a negative value, which indicates that the non-jamming approach cannot achieve secure maritime communications. Therefore, our approach can be adaptable to the extended scenario with multiple eavesdropping UAVs and jamming UAVs.

%
\section{Conclusion} 
\label{sec:conclusion}

\par This paper has implemented secure low-altitude maritime communications via UAV-assisted intelligent jamming. In the considered system, given the inherent trade-offs between conflicting objectives, we have formulated an SEMCMOP to jointly maximize the total secrecy rate of the system and minimize the total energy consumption of UAVs. To address the dynamic and long-term optimization problem, we have reformulated it into a POMDP. Then, we have proposed a GenAI-improved DRL algorithm, SAC-CVAE, which integrates a CVAE-based framework for policy disentanglement and optimization, as well as an LSTM-assisted prediction mechanism to enhance computational efficiency. Simulation results have shown that the UAV-assisted intelligent jamming approach significantly outperforms the non-jamming approach. Furthermore, comparison results have demonstrated that our proposed SAC-CVAE algorithm exhibits superior performance compared to other benchmark algorithms across various eavesdropper movement patterns, thereby maximizing the total secrecy rate while maintaining near-optimal total energy consumption of UAVs. Future research can explore extending this work to scenarios with imperfect marine user information as a promising direction.

\bibliographystyle{IEEEtran}
\bibliography{myref}{}

\vspace{-10 mm}
\begin{IEEEbiography}
[{\includegraphics[width=1in,height=1.25in,clip,keepaspectratio]{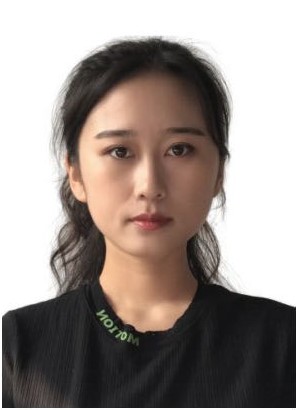}}]
{Jiawei Huang} received a BS degree in Software Engineering from Dalian Jiaotong University, and an MS degree in Software Engineering from Jilin University in 2019 and 2024, respectively. She is currently studying Computer Science at Jilin University to get a Ph.D. degree. Her current research interests are UAV networks and optimization.
\end{IEEEbiography}

\vspace{-10 mm}
\begin{IEEEbiography}
[{\includegraphics[width=1in,height=1.25in,clip,keepaspectratio]{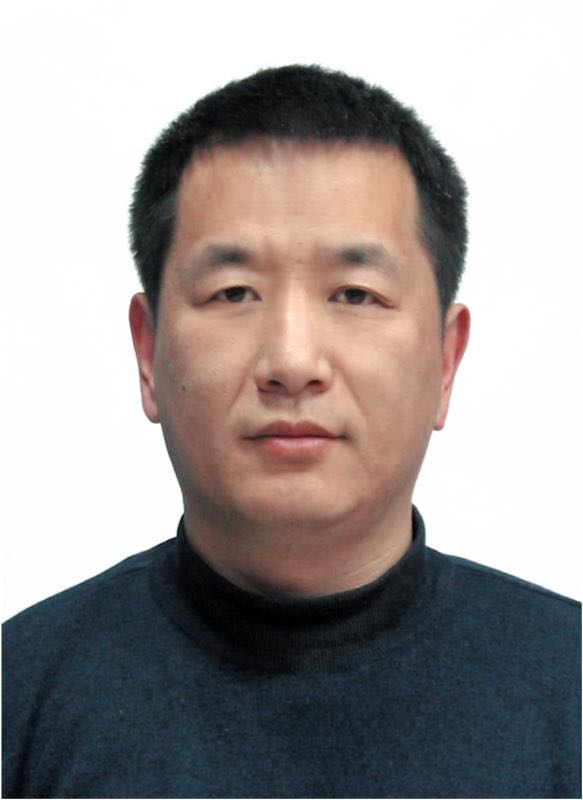}}]
{Aimin Wang} received the Ph.D. degree in Communication and Information System from Jilin University, Changchun, China, in 2004. He is currently a professor at Jilin University. His research interests are wireless sensor networks and QoS for multimedia transmission.
\end{IEEEbiography}

\vspace{-10 mm}
\begin{IEEEbiography}
[{\includegraphics[width=1in,height=1.25in,clip,keepaspectratio]{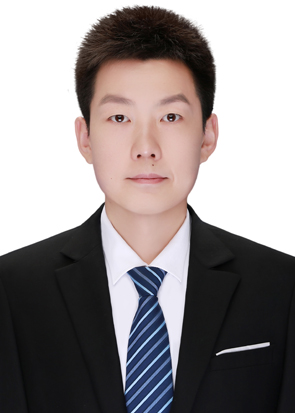}}]
{Geng Sun} (Senior Member, IEEE) received the B.S. degree in communication engineering from Dalian Polytechnic University, and the Ph.D. degree in computer science and technology from Jilin University, in 2011 and 2018, respectively. He was a Visiting Researcher with the School of Electrical and Computer Engineering, Georgia Institute of Technology, USA. He is a Professor in the College of Computer Science and Technology at Jilin University. Currently, he is working as a visiting scholar at the College of Computing and Data Science, Nanyang Technological University, Singapore. He has published over 100 high-quality papers, including IEEE TMC, IEEE JSAC, IEEE/ACM ToN, IEEE TWC, IEEE TCOM, IEEE TAP, IEEE IoT-J, IEEE TIM, IEEE INFOCOM, IEEE GLOBECOM, and IEEE ICC. He serves as the Associate Editors of IEEE Communications Surveys \& Tutorials, IEEE Transactions on Vehicular Technology, IEEE Transactions on Network Science and Engineering, and IEEE Networking Letters. He serves as the Lead Guest Editor of Special Issues for IEEE Transactions on Network Science and Engineering, IEEE Internet of Things Journal, IEEE Networking Letters. He also serves as the Guest Editor of Special Issues for IEEE Transactions on Services Computing, IEEE Communications Magazine, and IEEE Open Journal of the Communications Society. His research interests include UAV communications and networking, mobile edge computing (MEC), intelligent reflecting surface (IRS), generative AI, and deep reinforcement learning.
\end{IEEEbiography}

\begin{IEEEbiography}
[{\includegraphics[width=1in,height=1.25in,clip,keepaspectratio]{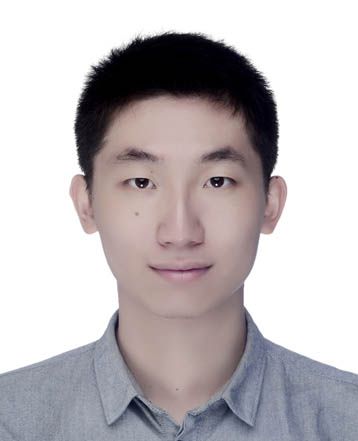}}]
{Jiahui Li} (Member, IEEE) received his B.S. in Software Engineering, and M.S. and Ph.D. in Computer Science and Technology from Jilin University, Changchun, China, in 2018, 2021, and 2024, respectively. He was a visiting Ph.D. student at the Singapore University of Technology and Design (SUTD). He currently serves as an assistant researcher in the College of Computer Science and Technology at Jilin University. His current research focuses on integrated air-ground networks, UAV networks, wireless energy transfer, and optimization.

\end{IEEEbiography}

\begin{IEEEbiography}
[{\includegraphics[width=1in,height=1.25in,clip,keepaspectratio]{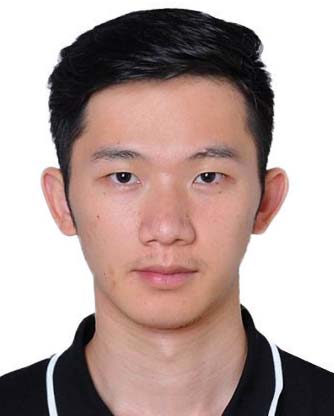}}]{Jiacheng Wang} received the Ph.D. degree from the School of Communication and Information Engineering, Chongqing University of Posts and Telecommunications, Chongqing, China. He is currently a Research Associate in computer science and engineering with Nanyang Technological University, Singapore. His research interests include wireless sensing, semantic communications, and metaverse.

\end{IEEEbiography}

\begin{IEEEbiography}[{\includegraphics[width=1in,height=1.25in,clip,keepaspectratio]{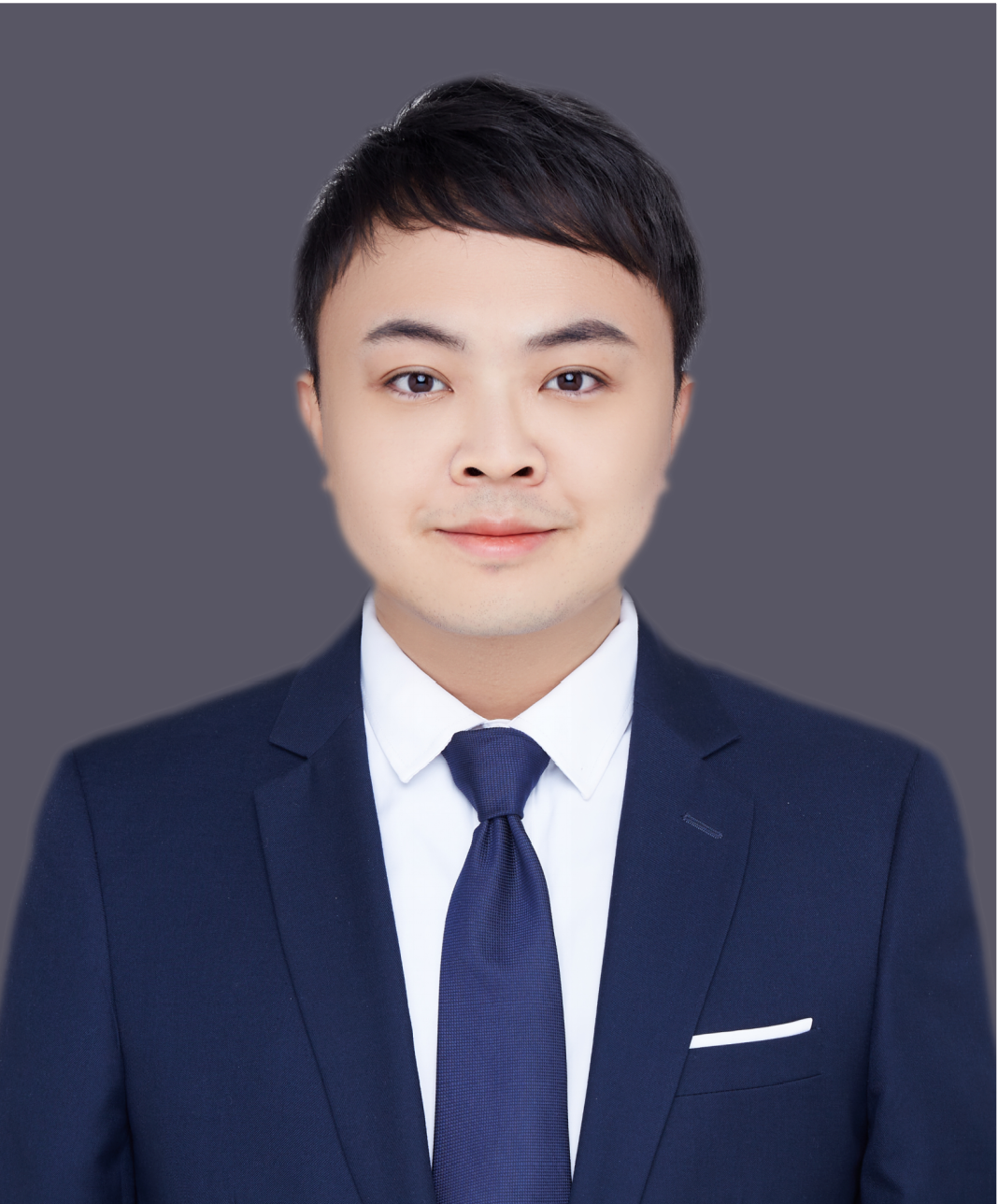}}]{Weijie Yuan} (Senior Member, IEEE) received the joint Ph.D. degree from the University of Technology Sydney, Ultimo, NSW, Australia, and Beijing Institute of Technology, Beijing, China, in 2019. In 2016, he was a Visiting Ph.D. Student with the Institute of Telecommunications, Vienna University of Technology, Austria. From 2017 to 2019, he was a Research Assistant with The University of Sydney, Visiting Associate Fellow with the University of Wollongong, and Visiting Fellow with the University of Southampton. From 2019 to 2021, he was a Research Associate with The University of New South Wales. He is currently the Series Lead Editor of IEEE Communications Magazine, and an Associate Editor for IEEE Transactions on Wireless Communications, IEEE Transactions on Green Communications and Networking, IEEE Communications Letters, IEEE Open Journal of Communications Society, and EURASIP Journal on Advances in Signal Processing. He is the Lead Editor of two feature topics in IEEE Communications Magazine.
\end{IEEEbiography}

\begin{IEEEbiography}
[{\includegraphics[width=1in,height=1.25in,clip,keepaspectratio]{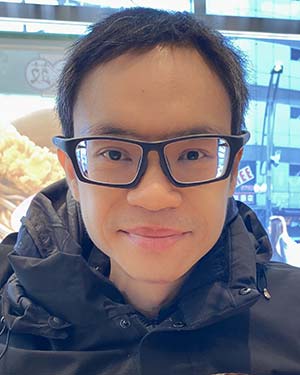}}]{Dusit Niyato} (Fellow, IEEE) received his Bachelor degree in Computer Engineering from King Mongkut's Institute of Technology Ladkrabang, Thailand, in 1999, Master and Ph.D. degrees from the University of Manitoba in 2005 and 2008, respectively. He is currently a professor in the College of Computing and Data Science, at Nanyang Technological University, Singapore. His research interests are in the areas of sustainability, edge intelligence, decentralized machine learning, and incentive mechanism design.
\end{IEEEbiography}

\vspace{-10 mm}
\begin{IEEEbiography}
[{\includegraphics[width=1in,height=1.25in,clip,keepaspectratio]{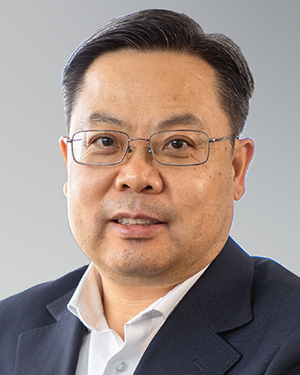}}] {Xianbin Wang} (Fellow, IEEE) received the Ph.D. degree in electrical and computer engineering from the National University of Singapore in 2001. He is a Professor and a Tier-1 Canada Research Chair of 5G and Wireless IoT Communications with Western University, Canada. He has over 600 highly cited journals and conference papers, in addition to over 30 granted and pending patents and several standard contributions. His current research interests include 5G/6G technologies, Internet of Things, machine learning, communications security, and intelligent communications. He is currently a member of the Senate, Senate Committee on Academic Policy, and Senate Committee on University Planning at Western University. He has been involved in many flagship conferences, including GLOBECOM, ICC, VTC, PIMRC, WCNC, CCECE, and ICNC, in different roles, such as General Chair, TPC Chair, Symposium Chair, Tutorial Instructor, Track Chair, Session Chair, and Keynote Speaker. He serves/has served as the Editor-in-Chief, an associate Editor-in-Chief, and an Editor/Associate Editor for over ten journals. He was the Chair of the IEEE ComSoc Signal Processing and Computing for Communications Technical Committee and is currently serving as the Central Area Chair of IEEE Canada. He is a Fellow of the Canadian Academy of Engineering and the Engineering Institute of Canada.
\end{IEEEbiography}
\end{document}